\documentclass[twocolumn]{aastex62}

\newcommand\teff{$T_{\mathrm{eff}}$}
\newcommand\logg{$\log g$}
\newcommand\vsini{$v \sin{i}$}

\newcommand\rp{$R_{\mathrm{p}}$}

\newcommand{\rearth}{$\mathrm{R_{\oplus}}$}

\newcommand{\kep}{\mbox{\textit{Kepler}}}
\newcommand{\gaia}{\mbox{\textit{Gaia}}}
\newcommand{\fearth}{$\mathrm{F_\oplus}$}
\newcommand{\slope}{$d \log R_{\mathrm{p}}$/$d \log M_\star$}

\newcommand{\mdot}{$\mathrm{M_\odot}$}
\newcommand{\rdot}{$\mathrm{R_\odot}$}

\newcommand{\rprstar}{$\frac{R_{\mathrm{p}}}{R_\star}$}

\newcommand{\nplanets}{\mbox{3898}}
\newcommand{\nhosts}{\mbox{2956}}

\usepackage{amsmath}
\usepackage{hyperref}
\usepackage{nameref}

\makeatletter
\newcommand{\labeltext}[2]{%
  \@bsphack
  \csname phantomsection\endcsname 
  \def\@currentlabel{#1}{\label{#2}}%
  \@esphack
}
\makeatother

\graphicspath{{./}{figures/}}

\shorttitle{Gaia-Kepler Catalog II: Exoplanets}
\shortauthors{Berger et al.}

\begin{document}

\title{The $Gaia$-$Kepler$ Stellar Properties Catalog. II. Planet Radius Demographics as a Function of Stellar Mass and Age}

\correspondingauthor{Travis Berger}
\email{taberger@hawaii.edu}

\author[0000-0002-2580-3614]{Travis A. Berger}
\affiliation{Institute for Astronomy, University of Hawai`i, 2680 Woodlawn Drive, Honolulu, HI 96822, USA}

\author[0000-0001-8832-4488]{Daniel Huber}
\affiliation{Institute for Astronomy, University of Hawai`i, 2680 Woodlawn Drive, Honolulu, HI 96822, USA}

\author[0000-0002-5258-6846]{Eric Gaidos}
\affiliation{Department of Earth Sciences, University of Hawai`i at M\={a}noa, Honolulu, HI 96822, USA}

\author[0000-0002-4284-8638]{Jennifer L. van Saders}
\affiliation{Institute for Astronomy, University of Hawai`i, 2680 Woodlawn Drive, Honolulu, HI 96822, USA}

\author[0000-0002-3725-3058]{Lauren M. Weiss}
\affiliation{Institute for Astronomy, University of Hawai`i, 2680 Woodlawn Drive, Honolulu, HI 96822, USA}

\begin{abstract}
\noindent
Studies of exoplanet demographics require large samples and precise constraints on exoplanet host stars. Using the homogeneous $Kepler$ stellar properties derived using $Gaia$ Data Release 2
by \cite{Berger2020a}, we re-compute $Kepler$ planet radii and incident fluxes and investigate their distributions with stellar mass and age. We measure the stellar mass dependence of the planet radius valley to be \slope\,=\,$0.26^{+0.21}_{-0.16}$, consistent with the slope predicted by \replaced{both photoevaporation}{a planet mass dependence on stellar mass} (0.24--0.35) and core-powered mass-loss (0.33). We also find first evidence of a stellar age dependence of the planet populations straddling the radius valley. Specifically, we determine that the fraction of super-Earths (1--1.8\,\rearth) to sub-Neptunes (1.8--3.5\,\rearth) increases from 0.61\,$\pm$\,0.09 at young ages ($<$\,1\,Gyr) to 1.00\,$\pm$\,0.10 at old ages ($>$\,1\,Gyr), consistent with the prediction by core-powered mass-loss that the mechanism shaping the radius valley operates over Gyr timescales. \added{Additionally, we find a tentative decrease in the radii of relatively cool ($F_{\mathrm{p}}$\,$<$\,150\,\fearth) sub-Neptunes over Gyr timescales, which suggests that these planets may possess H/He envelopes instead of higher mean molecular weight atmospheres}. We confirm the existence of planets within the hot \replaced{super-Earth}{sub-Neptunian} ``desert'' (2.2\,$<$\,$R_{\mathrm{p}}$\,$<$\,3.8\,\rearth, $F_{\mathrm{p}}$\,$>$\,650\,\fearth) and show that these planets are preferentially orbiting more evolved stars compared to other planets at similar incident fluxes. In addition, we identify candidates for cool ($F_{\mathrm{p}}$\,$<$\,20\,\fearth) inflated Jupiters, present a revised list of habitable zone candidates, and find that the ages of single- and multiple-transiting planet systems are statistically indistinguishable.
\vspace{1cm}
\end{abstract}

\section{Introduction} \label{sec:intro}

One of the most impactful exoplanet discoveries in recent years has been the planet radius ``valley'', a dip in the occurrence of \kep\ planets at $\approx$\,1.9\,\rearth\ separating super-Earth and sub-Neptune-sized exoplanets \citep{owen13,Fulton2017,Fulton2018}. The discovery of the radius valley was enabled by precise stellar parameters for subsamples of \kep\ host stars, such as those derived in the California-\kep\ Survey \citep[CKS,][]{Petigura2017,Johnson2017} and from asteroseismic constraints \citep{VanEylen2018}. More recently, \gaia\ parallaxes \citep{Brown2018,Lindegren2018} have better constrained the stellar radii of the vast majority of \kep\ host stars, followed by more detailed investigations of the valley as a function of stellar mass \citep{Fulton2018}, metallicity \citep{Owen2018}, planet orbital period, and stellar incident flux \citep{Berger2018c,Fulton2018}. Most recently, the radius valley has also been identified in the $K2$ sample \citep{Hardegree2020,Cloutier2020b}.

Several models have been proposed to explain the planet radius valley, including planet formation in a gas-poor disk \citep{Lee2014,Lee2016}, extreme ultraviolet (EUV) photoevaporation of planet atmospheres \citep{owen13,Owen2017,Owen2018,Lopez2018,Wu2019}, and core-powered mass-loss \citep{Ginzburg2016,Ginzburg2018,Gupta2019,Gupta2020}. Currently, photoevaporation and core-powered mass-loss are the two leading theories that can effectively explain the dependence of the gap on stellar mass, orbital period, and incident flux. However, observational studies have not yet been able to differentiate between these two theories. For example, to explain the radius valley as a function of stellar mass, photoevaporation requires a planet mass dependence on stellar mass \citep{Wu2019}. Multi-transiting systems hosting planets with mass measurements and radii both above and below the gap can distinguish between these models, but only a few examples are available \citep[e.g.,][]{Cloutier2020,Nowak2020}.

Stellar ages provide a new dimension to determine the physical mechanisms shaping exoplanet populations. Exoplanet properties are expected to change over time, such as a decrease in their radii from cooling and contraction \citep{Lopez2012} and atmosphere loss \citep{Ginzburg2016,Owen2017} or an increase in orbital eccentricity due to dynamical interactions between planets \citep{Weiss2018b}. However, ages are difficult to determine for stellar populations, such as \kep\ host stars, because available methods differ considerably across the Hertzprung-Russell (H-R) diagram. For instance, isochrone ages are effective on the upper main sequence ($M_\star$\,$\gtrsim$\,1\,\mdot), but are uninformative on the lower main sequence, where stellar rotation, activity, and lithium abundances provide more discriminatory power \citep{pont04,Epstein2014}. Asteroseismology provides precise stellar ages, but is generally only available for a small subset of mostly evolved exoplanet host stars \citep{Aguirre2015}.

So far, only a few studies have compared properties of exoplanets orbiting stars of different ages. \cite{Berger2018} found tentative evidence for the shrinking of planetary radii with stellar age based on the lithium abundances of CKS planet hosts differentiated by the Hyades 650\,Myr empirical lithium abundance (A(Li))--\teff\ curve \citep{Boesgaard2016}. Additionally, the Zodiacal Exoplanets In Time (ZEIT) survey yielded evidence for larger, younger planets in clusters where ages are already known \citep{Mann2018}. While both core-powered mass-loss and photoevaporation predict that the ratio of super-Earths to sub-Neptunes should increase over time, their timescales are very different:  photoevaporation acts on timescales of $\sim$\,100\,Myr \citep{Lopez2012,Owen2017} and core-powered mass-loss acts on timescales of $\sim$\,Gyr \citep{Gupta2019,Gupta2020}. Stellar ages are also critical to address other open questions in exoplanet radius demographics, such as the hot \replaced{super-Earth}{sub-Neptunian} desert \citep{Lundkvist2016,Berger2018c,dong18}, hot Jupiter inflation \citep{Guillot2002,Fortney2010,Baraffe2010,Baraffe2014,Laughlin2015,Laughlin2018,Komacek2020}, and the dynamical evolution of multiplanet systems \citep{Armitage2005,Spalding2016,RizzutoProp,Weiss2018b}.

Previous \kep\ stellar properties catalogs have not estimated ages due to inhomogeneous input parameters and the lack of precise parallaxes \citep{huber14,Mathur2017}. Here, we re-derive and analyze planet parameters using the updated stellar parameters by \cite{Berger2020a}\labeltext{B20}{B20} (\ref{B20} hereafter), the first homogeneous catalog of stellar \teff, \logg, radii, masses, densities, luminosities, and ages of \kep\ stars.

\section{Sample Selection and Methodology} \label{sec:methods}

\subsection{Host Star and Planet Sample} \label{sec:flagscuts}

First, we downloaded the KOI table on 10/13/19 from the NASA Exoplanet Archive, including 9564 planet candidates. Then, we cross-matched this table with our Table 2 in \ref{B20}, leaving 8875 planets. To avoid using stars with likely binary companions, we eliminated all stars with \gaia\ DR2 re-normalized unit-weight error (RUWE) $>$\,1.2 \citep[][\ref{B20}, and see Kraus et al. in prep]{Evans2018,Rizzuto2018,Bryson2020}. In addition, we discarded stars with unreliable isochrone-derived parameters (\texttt{iso\_gof}\,$<$\,0.99, \ref{B20}). We also removed all planets designated as false positives according to the \texttt{koi\_disposition} flag and those without reported planet-to-star radius ratios. We did not remove adaptive optics (AO)-detected binaries \citep{Furlan2017} to preserve number statistics, but we comment on their influence where relevant. Following these sample cuts, we retained \nhosts\ stars hosting \nplanets\ planets.

\subsection{Updated Planet Parameters} \label{sec:methodplanets}

We computed the updated planet radii utilizing the planet-to-star radius ratios provided in the KOI table from the NASA Exoplanet Archive and the stellar radii computed in \ref{B20}. In addition, we updated semimajor axes using the stellar masses in \ref{B20} and the orbital periods in \cite{Thompson2018}. Finally, we updated the incident fluxes for each planet by using the semimajor axes and the stellar luminosities from \ref{B20}. We tested the effect of our new stellar parameters on the planet-to-star radius ratios by computing new planet-to-star radius ratios using the transit period, duration, and depth values from \cite{Thompson2018}, the quadratic limb darkening coefficients from \cite{claret11}, Equation (9) from \cite{seager03}, and the small-planet-limit version of Equation (8) from \cite{mandel02}. We found that the differences between the \cite{Mathur2017}-derived and \ref{B20}-derived planet-to-star radius ratios were on the order of 3\% and within the 8\% median uncertainty in \cite{Thompson2018}. Therefore, these systematic effects are small and we neglect them here. We provide all of our planet parameters in Table \ref{tab:pars}.

\section{\kep\ Planet Host Stars}

\begin{figure}
\includegraphics[width=0.47\textwidth]{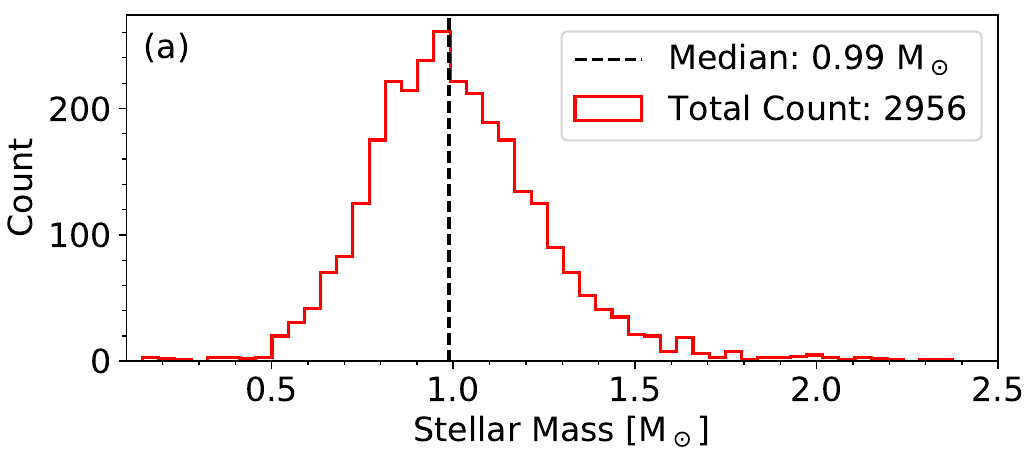}
\includegraphics[width=0.47\textwidth]{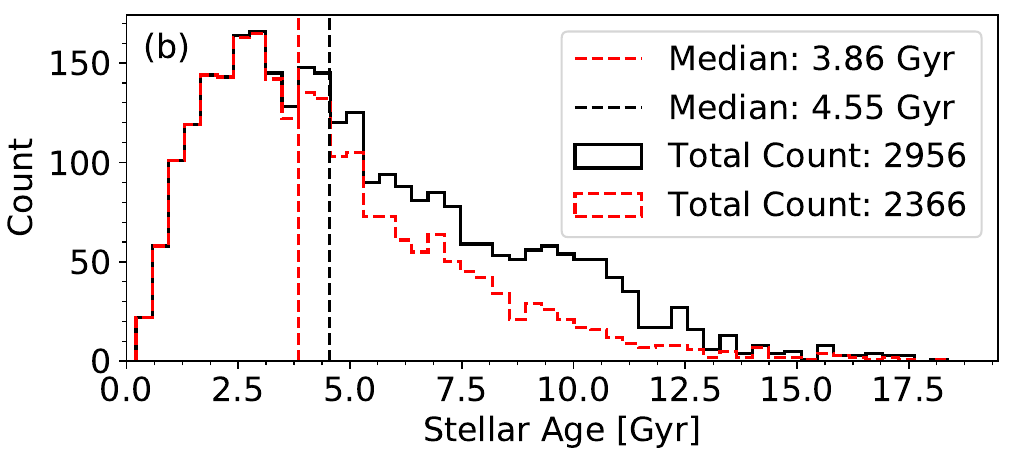}
\includegraphics[width=0.47\textwidth]{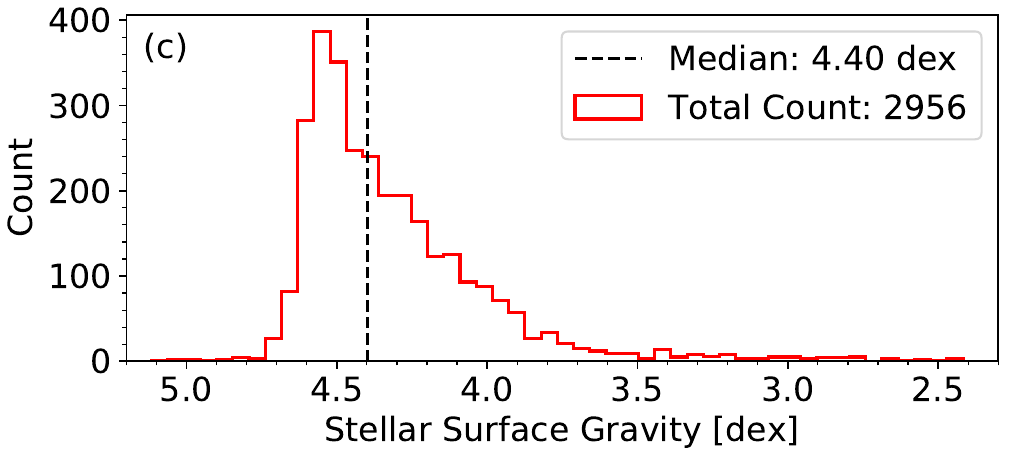}
\includegraphics[width=0.47\textwidth]{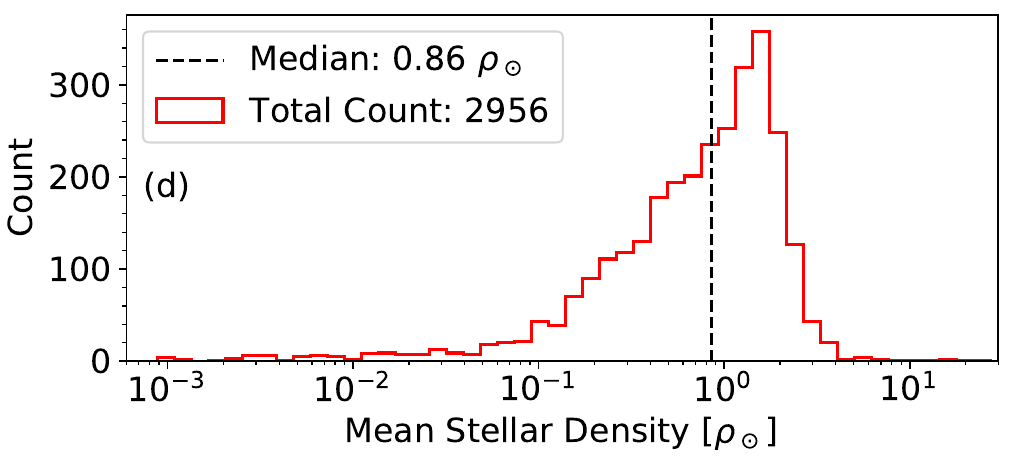}
\caption{Histograms of host star properties. The black dashed vertical lines illustrate the median value for each parameter. In Panel (b) we plot both the overall host sample (black) and those with reliable ages (red).} 
\label{fig:HostHist}
\end{figure}

\subsection{Histograms}

We plot histograms of physical parameters of the host star sample in Figure \ref{fig:HostHist}. Figure \ref{fig:HostHist}(a) shows the distribution of masses, which peak at solar mass, with a lower 1$\sigma$ bound of 0.78\,\mdot\ and an upper 1$\sigma$ bound of 1.19\,\mdot. The lowest and highest mass hosts are 0.14\,\mdot\ and 2.8\,\mdot, respectively.

Figure \ref{fig:HostHist}(b) displays the distributions of stellar ages, separated into the entire host sample (black) and an ``informative'' age sample (red). The informative age distribution ignores any hosts that have terminal age of the main sequence (TAMS) $>$\,20\,Gyr, which are stars that evolve too slowly in their main sequence lifetimes for isochrone-fitting to constrain their ages (\ref{B20}). The median ages are close to the age of the Sun, as expected from the \kep\ target selection \citep{batalha10}. Both distributions peak at 3\,Gyr with a tail towards old ages. Ages older than 14\,Gyr occur because the \ref{B20} model grid of MESA \citep{paxton11,paxton13,paxton15} Isochrones and Stellar Tracks \citep[MIST v1.2 with rotation,][]{choi16,dotter16} purposefully used an upper limit of 20\,Gyr to minimize grid edge-effects, which can bias the parameter posteriors. The two distributions in Figure \ref{fig:HostHist}(b) differ the most at $\approx$\,10\,Gyr, corresponding to M-dwarfs with uninformative ages producing flat posteriors with medians at half the age of the grid. The small number of hosts older than 14\,Gyr are stars whose input parameters place them on or close to the edge of the grid. Some of these old hosts are probably cool main sequence binaries similar to those identified in \cite{Berger2018c}, as stellar models are unable to reproduce their cool \teff\ and large radii at the age of the universe. Finally, we note that the overall age distribution is consistent with asteroseismic ages provided by the APOKASC2 catalog of \kep\ red giants \citep{Pinsonneault2018}.

Figures \ref{fig:HostHist}(c) and \ref{fig:HostHist}(d) shows the stellar surface gravity (\logg) and mean stellar density distributions of \kep\ host stars. These distributions are significantly different from the entire \kep\ stellar sample, which has another, smaller peak at \logg\,$\approx$\,2.5\,dex and $\rho_\star$\,$\approx$\,$10^{-3}$\,$\rho_\odot$. These peaks do not appear here because the percentage of giants with detected planets is much lower, given observational biases. Our median \logg\ and $\rho_\star$ are slightly smaller than solar, and the tails to smaller values are comprised of subgiants \citep{verner11,everett13,gaidos13,huber14}. There are a few giant hosts at the lowest \logg\ and density values, but many of these stars host unconfirmed, potential false positive planets \citep{sliski14}.

\subsection{The Host Star H-R Diagram}

\begin{figure*}
\resizebox{0.9\textwidth}{!}{\includegraphics{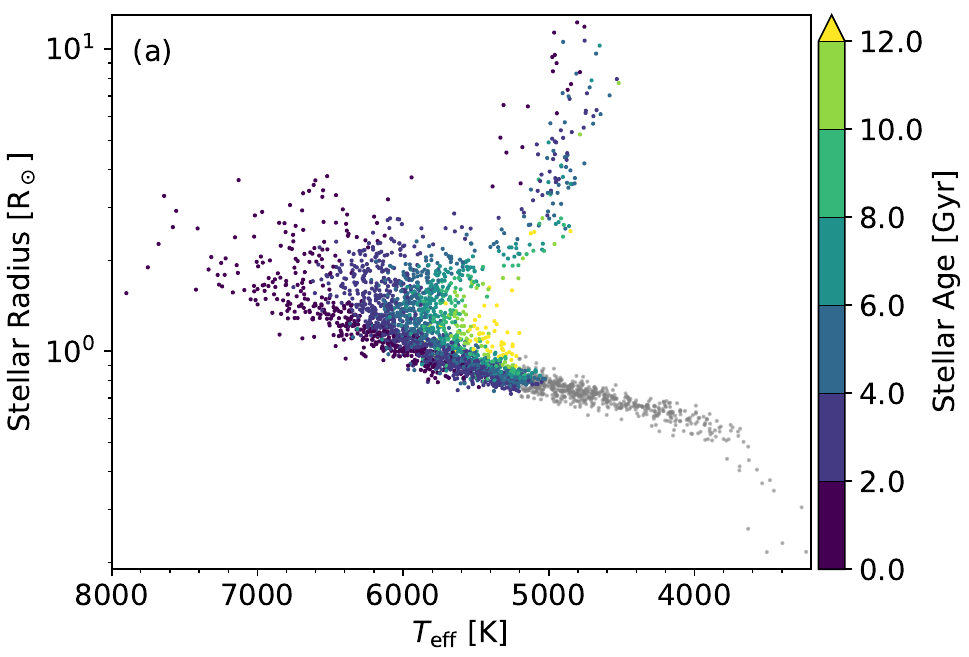}}
\resizebox{0.9\textwidth}{!}{\includegraphics{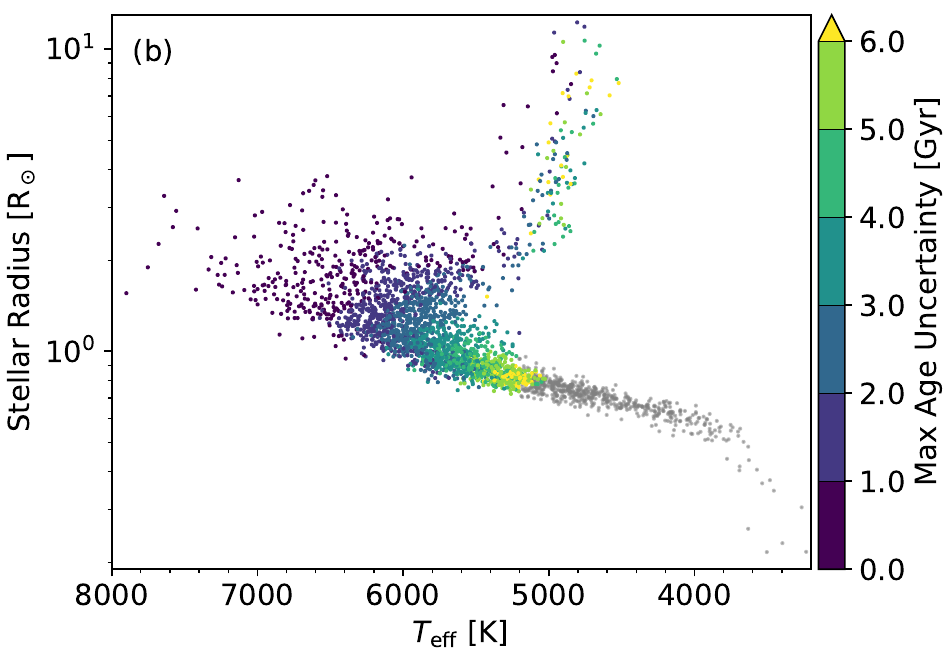}}
\centering
\caption{Hertzprung-Russell diagram of \kep\ planet host stars, colored by their isochrone age (top, colors capped at 12\,Gyr) and maximum absolute age uncertainties (bottom, colors capped at 6\,Gyr). The grey points have uninformative ages (TAMS\,$>$\,20\,Gyr) and/or low goodness-of-fit values. Nine stars hotter than 8000\,K are omitted from this plot.} \label{fig:HRHosts}
\end{figure*}

\begin{figure*}
\resizebox{\hsize}{!}{\includegraphics{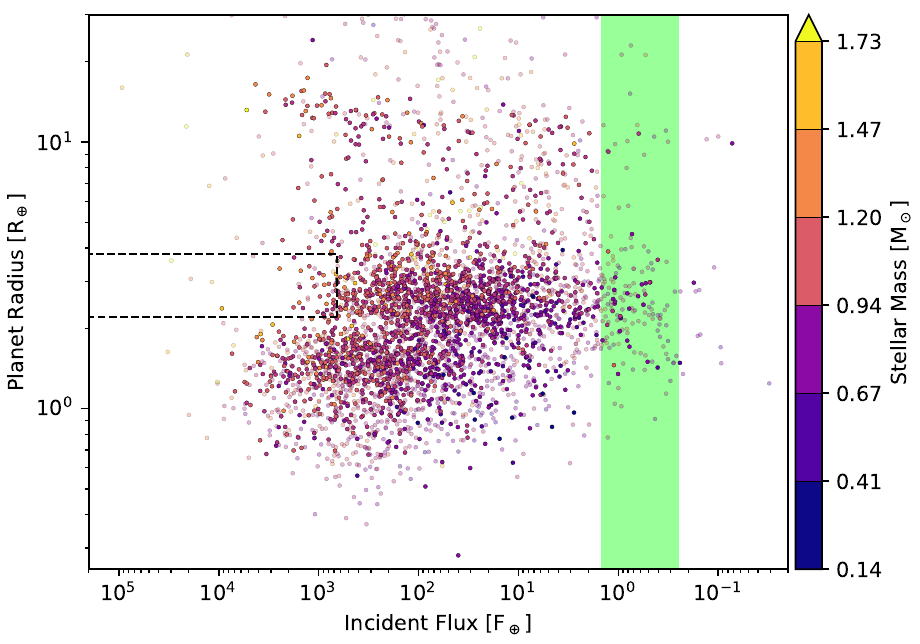}}
\caption{Planet radius versus incident flux for \kep\ exoplanets. Points are colored according to the host star mass as indicated by the color bar on the right. Planet candidates are shown as translucent points. The dashed line box shows the \replaced{super-Earth}{sub-Neptunian} desert identified in \cite{Lundkvist2016}, and the green bar indicates the approximate ``optimistic'' habitable zone defined by \cite{Kane2016}.} 
\label{fig:insol}
\end{figure*}

Figure \ref{fig:HRHosts} shows stellar radii versus temperature for the \kep\ host star sample, color-coded by age and maximum absolute age uncertainty. For the hottest stars the age values and uncertainties are less than 2\,Gyr. Both ages and age uncertainties increase smoothly towards cooler effective temperatures. Subgiants, compared to their main sequence counterparts, have smaller error bars due to rapid evolution on the subgiant branch, while their ages can vary significantly based on their \teff. On the giant branch, we also see that stars with the youngest ages and smallest age uncertainties are also the hottest and most massive, as more massive stars evolve more quickly than their lower mass counterparts. We note that giant ages are potentially unreliable because of the limitations of isochrone fitting with photometric colors and a solar neighborhood metallicity prior for stars that do not have spectroscopic measurements (see \ref{B20} for details).

The ages reach a maximum of $>$\,12\,Gyr at the main sequence turnoff for the least massive stars, while the maximum age uncertainties increase towards the lower main sequence, until they reach $>$\,6\,Gyr. Zero age main sequence (ZAMS) F stars have low age uncertainties between 0 and 2\,Gyr, while ZAMS G-dwarfs have moderate age uncertainties between 1 and 5 Gyr. For K-dwarfs, the ZAMS uncertainties are typically larger than 6\,Gyr. Some cool dwarfs with large/small radii have underestimated uncertainties due to grid edge effects. Finally, we see that all the late K--M-dwarfs have uninformative ages (TAMS\,$>$\,20\,Gyr, \ref{B20}). In particular, their observables provide limited information with which we can distinguish between the ages of these stars, which evolve slowly in the H-R diagram over 14\,Gyr.

Figure \ref{fig:HRHosts} illustrates variation in the effectiveness of isochrone placement for different types of stars, from subgiants (very effective) to K- and M-dwarfs (not effective at all). In addition, it demonstrates that the ages determined here can be used for the majority of \kep\ planet host stars, and hence \kep\ exoplanets, enabling one of the first investigations of how exoplanet properties change with stellar age. In the following investigations of stellar age, we ignore all grey points in Figure \ref{fig:HRHosts}.

\section{The Planet Radius Valley} \label{sec:pradvalley}

\subsection{Dependence on Stellar Mass} \label{sec:pradmass}

\begin{figure}
\resizebox{\hsize}{!}{\includegraphics{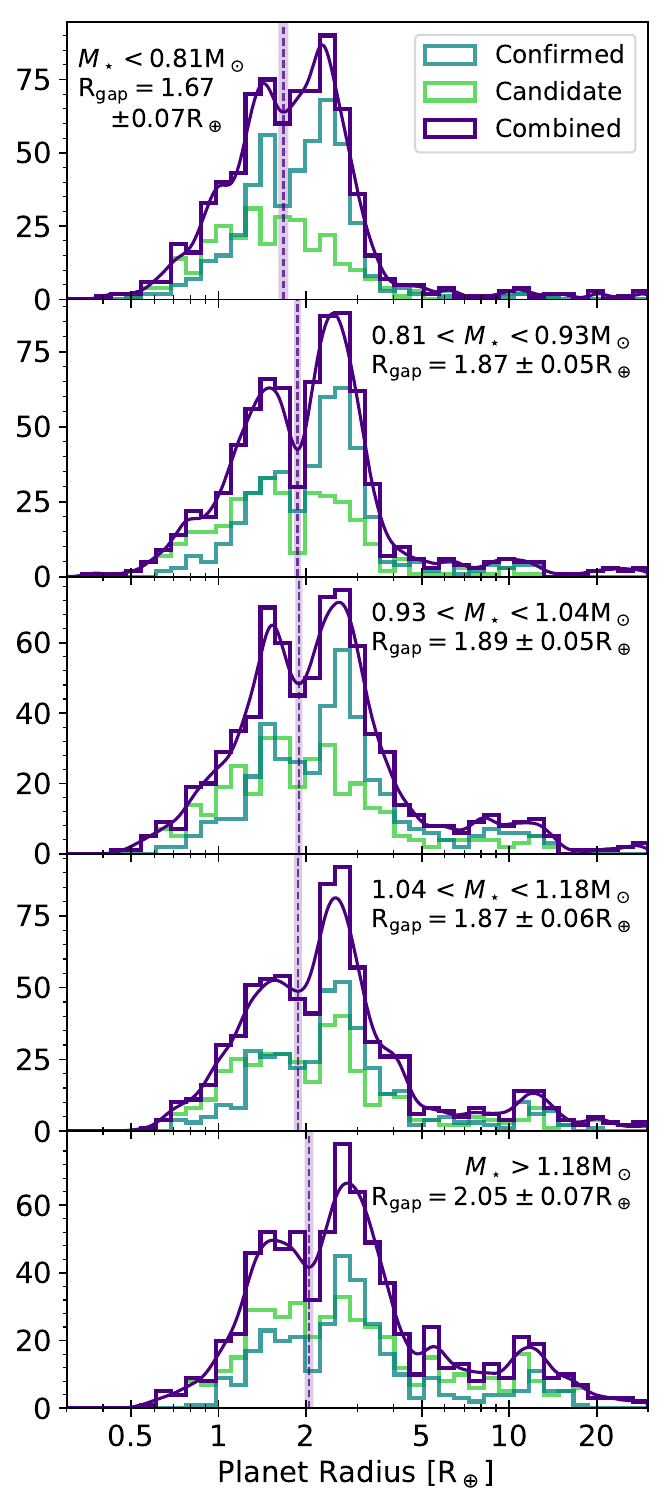}}
\caption{Distribution of \kep\ exoplanet radii, binned by stellar mass. Each panel is labeled by stellar mass and includes 767 planets. The teal and green histograms represent confirmed and candidate planets, respectively, while the purple histogram represents all planets. The purple, smooth lines show the 0.12\,$\log_{10}{\mathrm{R_\oplus}}$ bandwidth kernel density estimator (KDE) of the combined planet population for that panel. The vertical dashed purple lines and the shaded regions show the gap locations and their uncertainties from our Monte Carlo simulations.}
\label{fig:prad}
\end{figure}

\begin{figure*}
\resizebox{\hsize}{!}{\includegraphics{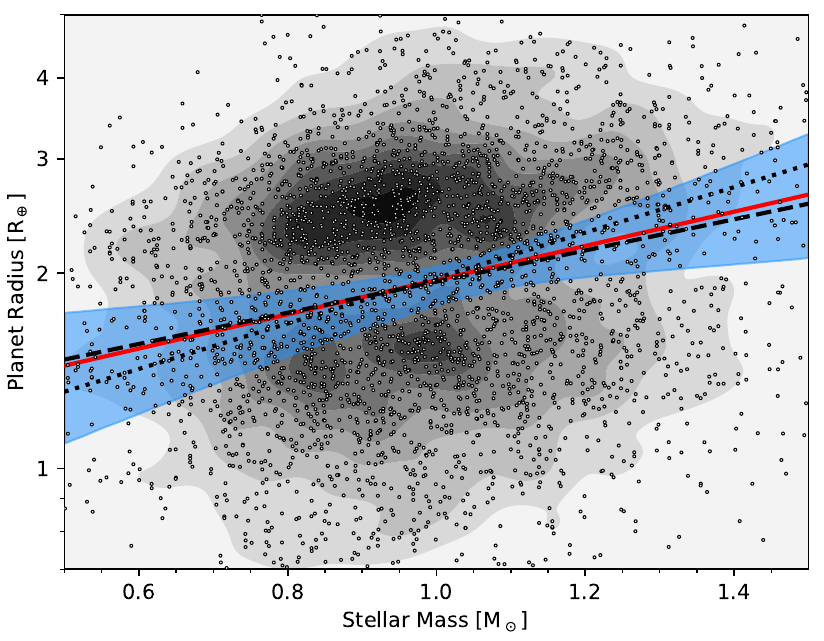}}
\caption{Planet radius versus stellar mass for \kep\ exoplanets. The contours represent the two-dimensional KDE distribution of the individual planets (small circles). Higher planet densities are darker colors. The blue shaded region illustrates the 1$\sigma$ bounds of our Monte Carlo and bootstrap simulations using \texttt{gapfit} \citep{Loyd2020}. The red line represents our best fit to the data, with a slope of \slope\,=\,$0.26^{+0.21}_{-0.16}$. The black \replaced{dashed line illustrates the lower bound predicted by photoevaporation \citep{Wu2019}, while the black dotted line is the largest slope predicted by photoevaporation (0.35)}{lines illustrate the slope assuming a relation between the planet mass and stellar mass of  $M_{\mathrm{p}}$\,$\propto$\,$M_\star$ (0.24, dashed line), and $M_{\mathrm{p}}$\,$\propto$\,$M_\star^{11/8}$ \citep[0.35, dotted line,][]{Wu2019}}.}
\label{fig:pradmass}
\end{figure*}

\subsubsection{Results} \label{sec:massres}

We will now use our revised planet parameters to address whether features in the planet radius distribution are dependent on host star properties, particularly stellar mass. Figure \ref{fig:insol} shows the distribution of planet radii as a function of incident flux, with individual planets colored according to host star mass. Planets at higher incident fluxes tend to orbit more massive hosts because these main sequence stars have higher luminosities than their less massive counterparts. Additionally, we observe a separation between super-Earths ($R_{\mathrm{p}}$\,$\lesssim$\,2\,\rearth) and sub-Neptunes ($R_{\mathrm{p}}$\,$\gtrsim$\,2\,\rearth) in a narrow region from ($\approx$\,10\,\fearth, $\approx$\,1.5\,\rearth) to ($\approx$\,1000\,\fearth, $\approx$\,2.2\,\rearth). This is the planet radius valley, a two-dimensional manifestation of the planet radius gap. The center of the gap is the planet radius at which the number density of planets is at a local minimum. Consistent with previous results \citep{VanEylen2018,Martinez2019}, we observe that the gap location increases in planet radius with increasing incident flux. Therefore, we also expect to find a dependence on stellar mass due to the stellar mass-luminosity relation.

Figure \ref{fig:prad} displays the planet radius distributions in five equally-populated stellar mass bins ranging from 0.14--2.8\,\mdot. We see a clear dependence of the planet radius gap's location in planet radius with stellar mass, increasing from 1.67\,\rearth\ at $M_\star$\,$<$\,0.81\,\mdot\ to 2.05\,\rearth\ at $M_\star$\,$>$\,1.18\,\mdot. To quantify this dependence, we computed the gap location in planet radius and its uncertainty by (1) drawing planet radii from normal distributions centered on the expected values and with standard deviation equal to their uncertainties and (2) computing the location of the gap by finding the relative minimum between 1--4\,\rearth\ in the kernel density estimate (KDE) distribution. If no gap was found, we repeated steps (1) and (2). We then computed the standard deviation for 100 successful gap samples to determine typical uncertainties in the gap location.

Figure \ref{fig:pradmass} directly compares planet radii with stellar mass. To quantify the stellar mass dependence of the planet radius gap's location in planet radius, we used \texttt{gapfit} \citep{Loyd2020} to fit a line of the form: 
\begin{equation}
    \log_{10}{R_{\mathrm{gap}}} = m*(\log_{10}{M_\star/\mathrm{M}_0}) + \log_{10}{\mathrm{R}_0},
\end{equation}
where the average gap depth is deepest in the 2D KDE distribution (red line). To ensure \texttt{gapfit} found the line corresponding to the deepest valley, we constrained the range of the \texttt{gapfit} search to 1.5--2.4\,\rearth.

To determine the uncertainty of the best-fit line while accounting for finite sampling, we drew each of the planet radii and stellar masses from normal distributions centered at each planet radius/stellar mass value with standard deviations given by the maximum of the upper and lower uncertainties for each planet radius and stellar mass. Next, we computed a 2D KDE from these simulated observations within the bounds of Figure \ref{fig:pradmass}, and constructed a sample of artificial planets drawn from the 2D KDE distribution. Then, we ran \texttt{gapfit} with $\mathrm{M_0}$\,=\,0.95\,\mdot, $\mathrm{R_{0,init}}$\,=\,1.89\,\rearth, $m_{init}$\,=\,0.27, and \texttt{sig}\,=\,0.15 (bandwidth in units of $\log_{10}{\mathrm{R_\oplus}}$ for y-axis and $\log_{10}{\mathrm{M_\odot}}$ for x-axis with no covariance) with 100 bootstraps to determine the best-fit slope for the valley in the newly sampled distribution of planet radii and stellar mass. We repeated this process for 100 re-draws of the planet radii and stellar masses and then computed uncertainties from these 100\,$\times$\,100 determinations of the best-fit line using the \texttt{uncertainty\_from\_boots} routine provided within \texttt{gapfit}. The blue shaded region in the slope in Figure \ref{fig:pradmass} represents the 1$\sigma$ uncertainty region of the best-fit line (\slope\,=\,0.26$^{+0.21}_{-0.16}$).

\added{Because core-powered mass-loss predicts that the positive slope of the radius valley is due to the stellar mass-luminosity relation \citep{Gupta2020}, we also compared the slopes of the valley across different stellar mass ranges. The slope of the mass-luminosity relation in log-log space, $\alpha$ \citep{Eker2018}, can be computed directly from the masses and luminosities of the host stars. We used \texttt{gapfit} with different but appropriate $\mathrm{M_{0}}$ and $\mathrm{R_{0,init}}$ to measure the slopes of the radius valley for stars of $\alpha$\,$\approx$\,2.9 ($M_\star$\,=\,0.14--0.72\,\mdot) and $\alpha$\,$\approx$\,5.5 ($M_\star$\,=\,0.81--1.05\,\mdot). We also compared the slopes of the radius valley for stellar mass ranges including $M_\star$\,$<$\,0.81\,\mdot\ and $M_\star$\,$>$\,1.18\,\mdot, 0.8\,$<$\,$M_\star$\,$<$1.0\,\mdot\ and 1.0\,$<$\,$M_\star$\,$<$\,1.2\,\mdot. However, we found that all of these 2D KDE distributions and their radius valley best-fit lines were very sensitive to the choice of the KDE bandwidth and initial parameters, and hence we do not report their results here.}

\subsubsection{Discussion}

Our results confirm the stellar mass dependence of the radius gap \citep{Fulton2018,Cloutier2020b}, with a best-fitting slope of \slope\ = $0.26^{+0.21}_{-0.16}$. \replaced{The uncertainty contains the range of slopes predicted by photoevaporation models \citep{Wu2019}}{This result is consistent with the range of slopes predicted by a dependence of planet mass on stellar mass \citep[which is required by photoevaporation models to explain the stellar mass dependence of the radius gap,][]{Wu2019},} and core-powered mass-loss models \citep[$\approx$\,0.33,][]{Gupta2020}. Therefore, we are unable to differentiate between these scenarios using the valley's slope in the planet radius-stellar mass diagram \citep[see also][]{Loyd2020}.

To evaluate whether our inability to differentiate between core-powered mass-loss and \replaced{photoevaporation}{a planet mass dependence on stellar mass} in planet radius-stellar mass space is simply a problem of sample size which might be ameliorated by future discoveries, we ran Monte Carlo simulations similar to those used in Figure \ref{fig:pradmass} to test various sample sizes and measurement precisions. We found that a sample size of 20,000 planets ($\approx$\,4 times the current number) assuming $typical$ planet radius and stellar mass errors of 1\% is needed to reduce the uncertainties in \slope\ to 0.04. This uncertainty would allow a 2--3$\sigma$ separation between the lower and upper bounds of \replaced{photoevaporation}{a planet mass dependence on stellar mass} \citep[0.24--0.35,][]{Wu2019}, which contain our measured value (0.26) and the value predicted by core-powered mass-loss \citep[0.33,][]{Gupta2020}. Even after combining all planets discovered by \kep, $K2$ \citep{Howell2014}, and those already and to be discovered by the Transiting Exoplanet Survey Satellite \citep[$TESS$,][]{ricker14}, both the sample size of $\sim$\,10,000 planets and the precision of our planet and stellar parameters would be insufficient to differentiate between photoevaporation\added{, which requires a planet-stellar mass dependence to explain the stellar mass dependence of the radius gap,} and core-powered mass-loss in planet radius-stellar mass space.

\subsection{Dependence on Stellar Age} \label{sec:pradage}

\subsubsection{Sample Selection} \label{sec:pradagesampsel}

\begin{figure}
\resizebox{\hsize}{!}{\includegraphics{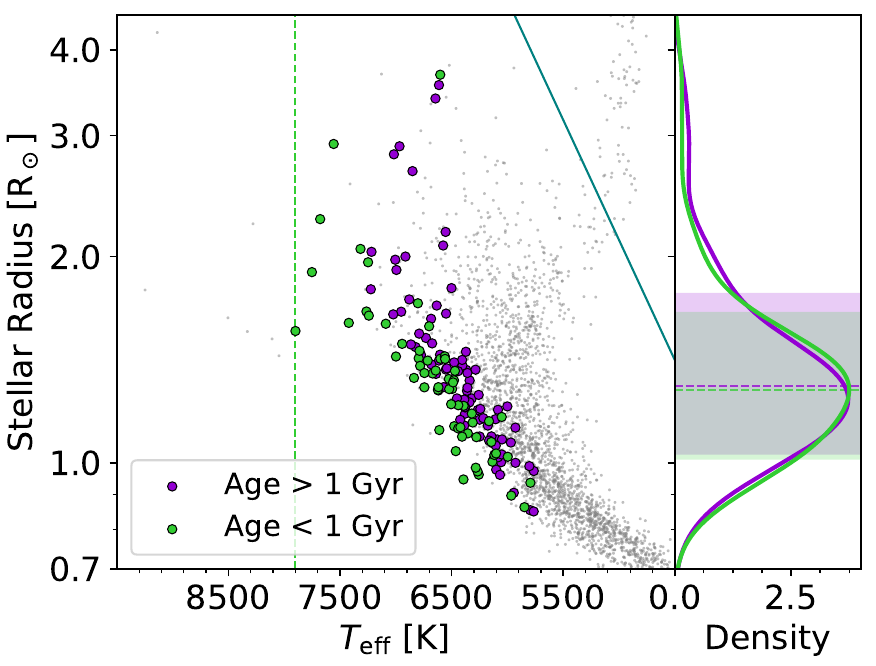}}
\caption{H-R diagram showing host stars colored according to their ages, as well as the marginalized distributions of stellar radii, with shaded areas representing the 16--84 percentile ranges. Purple host stars have ages greater than 1\,Gyr, and green host stars have ages younger than 1\,Gyr. The grey points are host stars that we do not include in our old and young samples: all evolved stars (above teal line), stars hotter than 7900\,K (left of the dashed, vertical green line), and old stars with dissimilar mass, radius, and/or metallicity to the young stellar sample.}
\label{fig:stellaroldyoung}
\end{figure}

\begin{figure*}
\resizebox{\hsize}{!}{\includegraphics{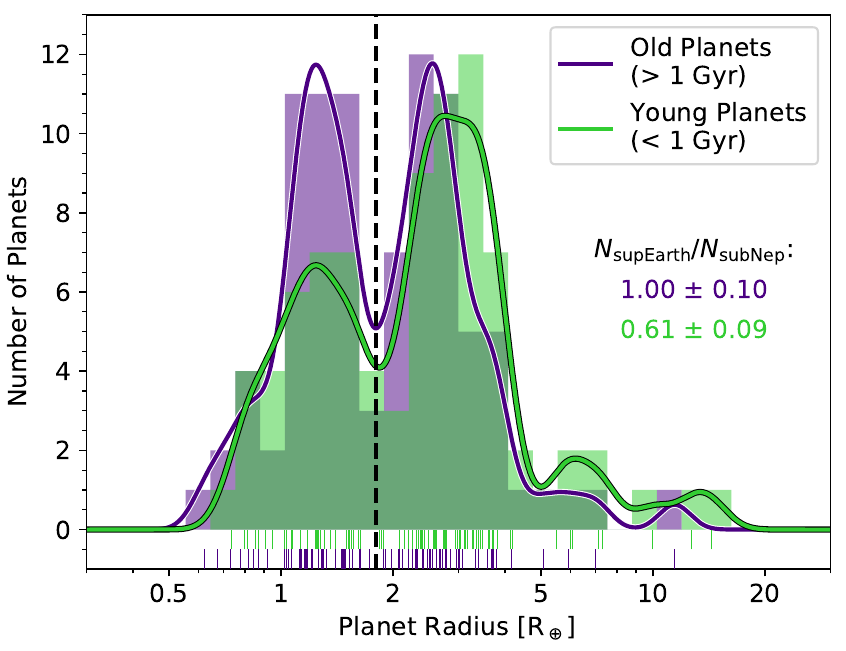}}
\caption{Radius distributions in KDE and histogram form of \kep\ exoplanets with ages younger than (green) or older than (purple) 1\,Gyr. The individual planet radii are plotted below as vertical ticks. The black dashed vertical line at 1.8\,\rearth\ is the \cite{Fulton2017} gap radius, which separates super-Earths and sub-Neptunes. The ratio of super-Earths (1--1.8\,\rearth) to sub-Neptunes (1.8--3.5\,\rearth) significantly increases from young ages (0.61\,$\pm$\,0.09) to old ages (1.00\,$\pm$\,0.10).}
\label{fig:pradoldyoung}
\end{figure*}

\begin{figure}
\resizebox{\hsize}{!}{\includegraphics{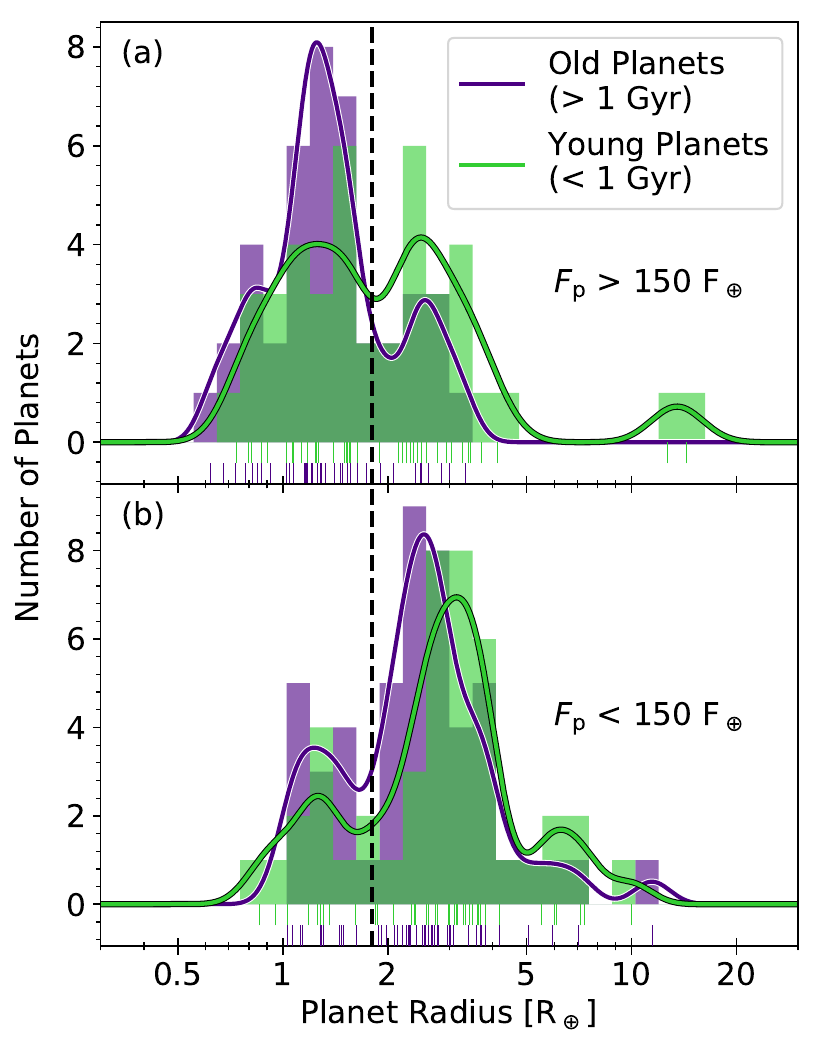}}
\caption{Same as Figure \ref{fig:pradoldyoung} but separating the samples into high ($>$\,150\,\fearth) and low ($<$\,150\,\fearth) incident flux.}
\label{fig:pradoldyounghighlow}
\end{figure}

The photoevaporation and core-powered mass-loss mechanisms operate on different timescales and consequently predict different dependence of the planet radius distribution with age. Thus, we grouped planets according to the age of their host star. To ensure we only included planets with reliable properties, we removed 64 planets with radii greater than 30\,\rearth. Additionally, we removed 97 planets with grazing transits defined as \rprstar\ + $b$ $>$ 1, where $b$, the impact parameter, is how far away from the center of the star's disk the center of the planet transits at mid-transit (0 being at the center, and 1 being at the limb). We used $\log_{10} \frac{R_\star}{\mathrm{R_\odot}} < 0.00035(T_{\mathrm{eff}}-4500) + 0.15$ \citep[similar to Equation 1 in][]{Fulton2017} to remove giant hosts, which have a high false-positive rate \citep{sliski14}. We also removed 1081 planets orbiting hosts without spectroscopic metallicity constraints because isochrone age estimates are metallicity-sensitive \citep{Howes2019}. After this metallicity cut, 2545 planets remained.

We selected 1\,Gyr as our separator between young and old systems because core-powered mass-loss acts on $\sim$\,Gyr timescales \citep{Gupta2020}. We could not make an age cut at 100\,Myr, the timescale relevant for photoevaporation \citep{Owen2017}, because we have no stars with isochrone ages $<$\,100\,Myr. Likewise, a cut at the age of the Hyades \citep[$\approx$\,650\,Myr,][]{Boesgaard2016} includes too few young planets for a robust statistical comparison. We removed seven young planets with host effective temperatures hotter than 7900\,K because their \teff\ and masses were outliers compared to the rest of the age sample. Eighty-five planets orbit hosts with median posterior ages younger than 1\,Gyr, while the remaining 2453 planets orbit hosts with median posterior ages older than 1\,Gyr.

To remove degeneracies between stellar mass, age, and metallicity in the old and young planet samples, we used the \texttt{NearestNeighbors} function in \texttt{scikit-learn} \citep{scikit-learn} to choose the two nearest old neighbors for every young host in stellar mass, radius, and metallicity. Because our matching function occasionally chose old neighbors such that multiple young stars are matched to the same old star, we removed any duplicate old hosts to avoid counting the same old host twice. We chose to use the two nearest neighbors instead of either one or three because the former selected fewer old planets than young planets after dropping duplicates, while the latter produced inferior stellar property-matched samples, especially in stellar mass. Our resulting property-matched sample included 90 old planets. In addition to removing stellar population biases, this careful sample selection also reduced potential detection biases for small planets between the old and young samples, all while retaining the full young planet sample. We used K-S tests to compare the stellar mass, radius, and metallicity distributions of the old and young samples, producing p-values of 0.24, 0.58, and 0.997, respectively, confirming the distributions are statistically similar. Figure \ref{fig:stellaroldyoung} shows an H-R diagram of the host star sample after making the above cuts, separated by the relevant age bins. We flag these planets as ``Old'' and ``Young'' in Table \ref{tab:pars}.

\subsubsection{Results}

Figure \ref{fig:pradoldyoung} shows the planet radius distributions of the old and young planet samples. We observe that the gap occurs at approximately the same planet radius in both the old and young planet distributions. Remarkably, we also observe that the ratio of super-Earths compared to sub-Neptunes significantly increases from young to old stellar ages.

To quantify this age dependence, we computed the uncertainties in the ratio of super-Earths to sub-Neptunes using Monte Carlo simulations to draw each old and young planet radius from a normal distribution given its measured value and uncertainty. We define super-Earths as planets between 1--1.8\,\rearth, and sub-Neptunes as planets between 1.8--3.5\,\rearth\ \citep{Fulton2017}. We then counted the number of super-Earths and sub-Neptunes, repeated this process 1000 times for both the old and the young distributions and then computed the standard deviation of the ratios of super-Earths to sub-Neptunes.

Our data show a significant increase in the fraction of super-Earths as a function of age, with the fraction of super-Earths to sub-Neptunes increasing from 0.61\,$\pm$\,0.09 at young ages ($<$\,1\,Gyr) to 1.00\,$\pm$\,0.10 at old ages ($>$\,1\,Gyr). This result is insensitive to the choice of impact parameter cut ($\gtrsim$\,2$\sigma$ for $b$\,$<$\,0.7--0.9), gap location ($\gtrsim$\,3$\sigma$ for 1.9--2.0\,\rearth), and radius range used to define super-Earths and sub-Neptunes ($\gtrsim$\,4$\sigma$ for 0.8--1.8\,\rearth\ and 1.8--5\,\rearth, respectively). Similarly, if we instead used 1533 old planets hosted by old stars larger than 0.9\,\rdot\ rather than the property-matched sample described in the previous section, we computed a $\gtrsim$\,3$\sigma$ difference in the ratios of young and old super-Earths to sub-Neptunes. Reassuringly, \cite{Fulton2017} computes the $occurrence$ ratio of super-Earths to sub-Neptunes for the entire planet sample to be 0.8\,$\pm$\,0.2, approximately the average of our old and young ratios.

We also compared the low ($F_{\mathrm{p}}$\,$<$\,150\,\fearth) and high ($F_{\mathrm{p}}$\,$>$\,150\,\fearth) flux planet radius distributions for old and young exoplanets (Figure \ref{fig:pradoldyounghighlow}). We chose 150\,\fearth\ because it splits the young sample of planets almost in half: 41 young and 41 old planets receive more than 150\,\fearth\ and 44 young and 49 old planets receive less than 150\,\fearth. Interestingly, we observe stark differences between old and young planets. At high incident flux (Figure \ref{fig:pradoldyounghighlow}(a)), we observe a large difference between the young and old planet radius distributions. We compute the ratios of super-Earths to sub-Neptunes to be 1.00 and 2.67 for the young and old planets, respectively. However, at low incident flux (Figure \ref{fig:pradoldyounghighlow}(b)), the overall distributions do not show a strong difference as a function of age, with tentative evidence that old sub-Neptunes are smaller than young sub-Neptunes. We ran K-S tests to quantitatively compare the old and young distributions in both panels, and found they were statistically distinguishable in the top panel (p-value\,=\,0.02) and indistinguishable in the bottom panel (p-value\,=\,0.11).

\begin{figure}
\resizebox{\hsize}{!}{\includegraphics{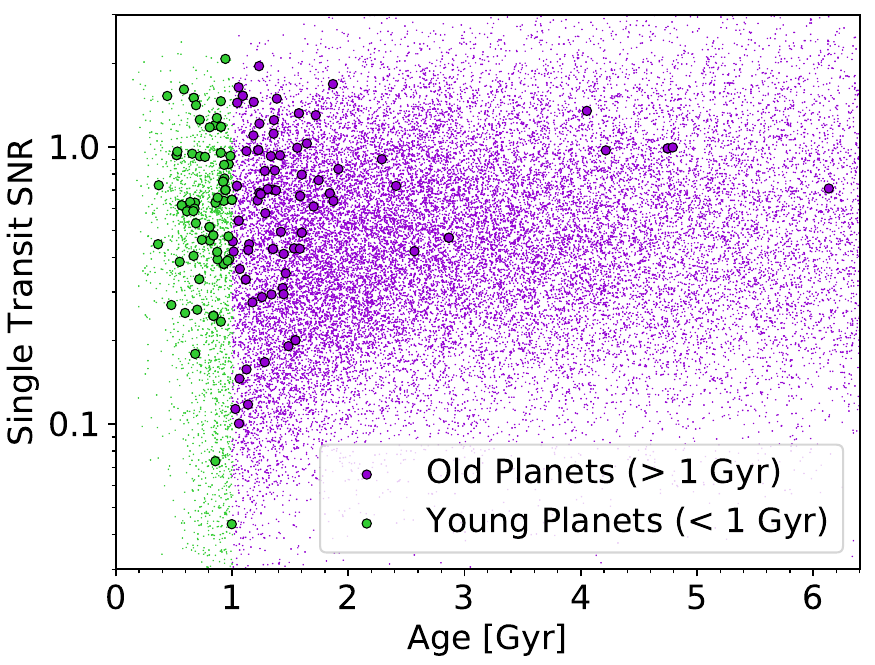}}
\caption{Single transit signal-to-noise ratios (SNR) for an Earth-size planet with a three-hour transit duration orbiting each individual planet host used in Figures \ref{fig:stellaroldyoung}--\ref{fig:pradoldyounghighlow} (large circles) and \kep\ target star (dots). Larger values on this plot indicate that planets are easier to detect. We use the CDPP3 values from \cite{Christiansen2012} and Equation B2 in \cite{Petigura2018} to compute the SNR.}
\label{fig:singtrans}
\end{figure}

A potential bias in our results is the sensitivity of planet detection to stellar age. This is because young stars are typically noisier due to their increased activity \citep{Skumanich1972} and thus smaller planets might not be detected around them. To test this, we evaluated the CDPP3 values \citep{Christiansen2012} to determine the single transit signal-to-noise ratio (SNR) for an Earth-size planet with a three hour transit duration \citep[as in][]{Petigura2018} for each host star in Figure \ref{fig:singtrans}. This comparison also accounts for differences in the stellar radius distributions, although we control for these differences in our sample selection.

We find that the young planets typically have similar SNR compared to the old planets, and hence are just as difficult to detect. The large scatter of points in Figure \ref{fig:singtrans} dominate what would be small differences in the old and young median single transit SNR values. Similarly, when we compare the median single transit SNR values the overall \kep\ sample with cuts similar to those in Figure \ref{fig:stellaroldyoung}, we find no significant difference between the old median SNR ($\approx$\,0.43) and young median SNR ($\approx$\,0.40), given the large point-by-point scatter. Therefore, we conclude that our results are robust against a planet detection bias with stellar age.

Another bias that could affect our results is the presence of undetected binary companions, which would cause stars to be incorrectly assigned old ages. While we have removed most wide binaries using \gaia\ and de-biased stellar photometry for binary contamination (\ref{B20}), it is inevitable that some close binaries will contaminate our old and young samples. After removing all AO-detected companions \citep{Furlan2017}, the young and old distributions in Figure \ref{fig:pradoldyoung} remain mostly unchanged. However, the distributions in Figure \ref{fig:pradoldyounghighlow}(a) are sensitive to the removal of AO-detected binaries, as binaries make stars appear more luminous and increase the apparent irradiance of the planets.

To test the potential for binaries to bias our results, we estimated the number of planets that may have been mistakenly placed in the older sample because of binary contamination. We adopted a typical binary mass ratio ($M_{\mathrm{prim}}/M_{\mathrm{sec}}$\,$\approx$\,0.5) and binary fraction for solar-type stars ($F_{\mathrm{bin}}$\,$\approx$\,40\%) determined in \cite{Raghavan2010} and \cite{Moe2017}. We also used the +50\% isochrone age bias determined from the 1.15\,\mdot\ star in Figure 4 of \ref{B20}. Using these assumptions, we predict that 55\,planets\,$\times$\,$F_{\mathrm{bin}}$\,$\approx$\,22 planets are orbiting stars with isochrone ages between 1--1.5\,Gyr that should be younger than 1\,Gyr due to binary contamination. We then shifted these 22 planets into the young distribution assuming (1) they mimic the old distribution's ratio of super-Earths to sub-Neptunes and (2) their measured radii are not affected significantly by binary contamination. Consequently, we computed a 0.76\,$\pm$\,0.09 ratio of super-Earths to sub-Neptunes for the young planet distribution. Comparing this to the 1.00\,$\pm$\,0.10 ratio of super-Earths to sub-Neptunes for the old planet distribution, we still arrive at a $>$\,2$\sigma$ difference between the old and young distributions. Given our conservative assumptions, we conclude that undetected stellar companions will not significantly affect our results on the age dependence of the radius valley.

\subsubsection{Discussion}

Figure \ref{fig:pradoldyoung} suggests that sub-Neptunes evolve to become super-Earths over Gyr timescales. This is consistent with the core-powered mass-loss mechanism \citep{Gupta2020}, which predicts that the transition of sub-Neptunes to super-Earths occurs over Gyr timescales as the planets gradually lose their atmospheric envelopes due to their heated cores. While our result supports the core-powered mass-loss mechanism, it does not rule out the possibility that photoevaporation is also acting on the observed planet population. Given that photoevaporation is expected to occur in the first $\sim$\,100\,Myr \citep{Owen2017,Wu2019} of a planet's lifetime and our 100\,Myr minimum grid age (\ref{B20}), we are mostly insensitive to any planetary evolution before 100\,Myr ages, barring extreme planet radius evolution during that time. Our Monte Carlo simulations do not account for the uncertainties in the stellar ages and hence it is likely that there is contamination between the old and young planets. However, given the magnitude of the errors and the difference in the number of old and young stars in our sample, it is more likely that older stars are contaminating the younger bins (Eddington bias). If this bias truly exists in our sample, any observed differences between the old and young samples will be reduced by this bias, thus the true difference between young and old planets is greater.

Our result agrees with \cite{Berger2018}, who used lithium abundances relative to the Hyades to separate young (A(Li)$_\star$\,$>$\,A(Li)$_{\mathrm{Hyades}}$) and old (A(Li)$_\star$\,$<$ A(Li)$_{\mathrm{Hyades}}$) planets. In particular, \cite{Berger2018} find that there is a significant difference in the sizes of planets in the old and young planet radius histograms and that young planets are larger. This result was tentative ($\approx$\,2--3$\sigma$) due to the small sample size of the young planets as compared to the old planets. Although \cite{Berger2018} does not explicitly quantify the number of super-Earths and sub-Neptunes, the largest difference between their old and young samples occurs at $\gtrsim$\,2\,\rearth, where the number of sub-Neptunes in the young sample is significantly greater than the number sub-Neptunes in the old sample.

Our observation of a similar gap radius for young and old planets is expected, as we carefully chose stellar samples with similar stellar mass, which should produce a gap at a similar location in the distribution of planet radii (\S\ref{sec:pradmass}). While core-powered mass-loss predicts that the gap's location should increase slightly to larger radii in the first $\sim$\,Gyr \citep{Gupta2020}, such behavior will be hard to definitively establish given typical uncertainties on stellar age, mass, and planet radius. Meanwhile, photoevaporation does not predict significant gap movement after the first $\sim$\,100\,Myr \citep{owen16,Owen2017}.

We interpret that the differences between the old and young distributions as a function of incident flux in Figure \ref{fig:pradoldyounghighlow} are consistent with core-powered mass-loss, which predicts that planets that receive higher incident fluxes may experience increased and potential runaway mass-loss over timescales of $\sim$\,Gyr due to increased equilibrium temperatures at the Bondi radius \citep{Ginzburg2016,Ginzburg2018,Gupta2019,Gupta2020}. Conversely, sub-Neptunes that receive low incident fluxes (Figure \ref{fig:pradoldyounghighlow}(b)) will simply cool and contract, shifting from the larger radii ($\approx$\,3\,\rearth) at young ages to the smaller radii ($\approx$\,2.5\,\rearth) at old ages. This is predicted by photoevaporation as well \citep{Lopez2012,Owen2017}, but it cannot describe the difference in high incident flux distributions over Gyr timescales. \added{In addition, the marginally significant difference (p-value = 0.11) in the sizes of old and young sub-Neptunes at low incident fluxes may suggest that these planets have significant H/He envelopes instead of higher mean molecular weight envelopes that cannot produce the same magnitude of contraction \citep[][but see also \citeauthor{Howe2015} \citeyear{Howe2015}]{Nettelmann2011,Lopez2012,Lopez2014,Lopez2017}. Alternatively, the shrinking of these planets could also be caused by H$_2$/He ingassing at orbital periods $<$\,100\,days \citep{Kite2020}. More theoretical and observational work is required to evaluate the composition of these atmospheres \citep{Owen2019} and} we caution that these inferences are tentative, especially because of the small number of planets contained in \replaced{the young low flux (44 planets) and high flux (41 planets) distributions}{the old and young low flux (49 and 44 planets, respectively) and old and young high flux (41 and 41 planets, respectively) distributions}. Additionally, both the young and the old low flux planet radius distributions are likely biased more significantly by selection effects than the high flux distributions, as small planets at low incident fluxes are more difficult to detect.

\subsection{Planets Within the Gap} \label{sec:ingap}

According to photoevaporation, which produces radius changes within the first 100\,Myr of a host's lifetime \citep{Owen2017}, we should not see any planets within the gap \citep{VanEylen2018} at old ages. Conversely, the $\sim$\,Gyr timescales of core-powered mass-loss \citep{Gupta2020} suggest that we should find a few planets in the gap region as they transition from sub-Neptunes to super-Earths. It is currently unclear whether any planets firmly exist within the radius valley. Large population studies have revealed that there are at least a few planets that fall within the planet radius gap \citep{Berger2018c,Fulton2018}, although smaller, more precise planet samples have revealed a complete lack of planets within the gap \citep{VanEylen2018}.

To investigate this, we used \texttt{gapfit} \citep{Loyd2020} to determine the best-fit parameters in the planet radius-incident flux diagram using:
\begin{equation}
    \log_{10}{R_{\mathrm{gap}}} = m*(\log_{10}{F_{\mathrm{p}}/\mathrm{F}_0}) + \log_{10}{\mathrm{R}_0},
\end{equation}
assuming a pivot point $\mathrm{F}_0$\,=\,100\,\fearth\ and the optimal Gaussian kernel width \texttt{sig}\,=\,0.15 \citep[\texttt{gapfit,}][]{Loyd2020}. We found $m$\,=\,$d\log R_{\mathrm{p}}$/$d\log F_{\mathrm{p}}$\,=\,0.057 and $\mathrm{R}_0$\,=\,1.86\,\rearth. Next, we computed parallel lines by varying the $\mathrm{R}_0$ parameter by typical uncertainties ($\pm$\,0.09\,\rearth) determined from a combination of Monte Carlo and bootstrap simulations, ignoring uncertainties in the slope. We then isolated all confirmed planets that were within the log-log lines with slopes, $m$\,=\,$d\log R_{\mathrm{p}}$/$d\log F_{\mathrm{p}}$\,=\,0.057, and pivot point central gap radii, $\mathrm{R}_0$\,=\,1.77 and 1.95\,\rearth. We also removed all planets with 1$\sigma$ errors in planet radius that would place them outside the log-log lines representing bounds of the planet radius gap as a function of incident flux.

Following these cuts, five planets remain in the gap:  \kep-11\,b, \kep-110\,b, \kep-114\,c, \kep-634\,b, and \kep-887\,b. We conclude that these five planets may currently undergo core-powered mass-loss, but caution that they are only $\sim$\,1$\sigma$ removed from the gap boundaries. Additional follow-up observations will be required to definitively identify planets inside the radius valley.

\section{The Hot Sub-Neptunian Desert} \label{sec:desert}

\begin{figure}
\resizebox{\hsize}{!}{\includegraphics{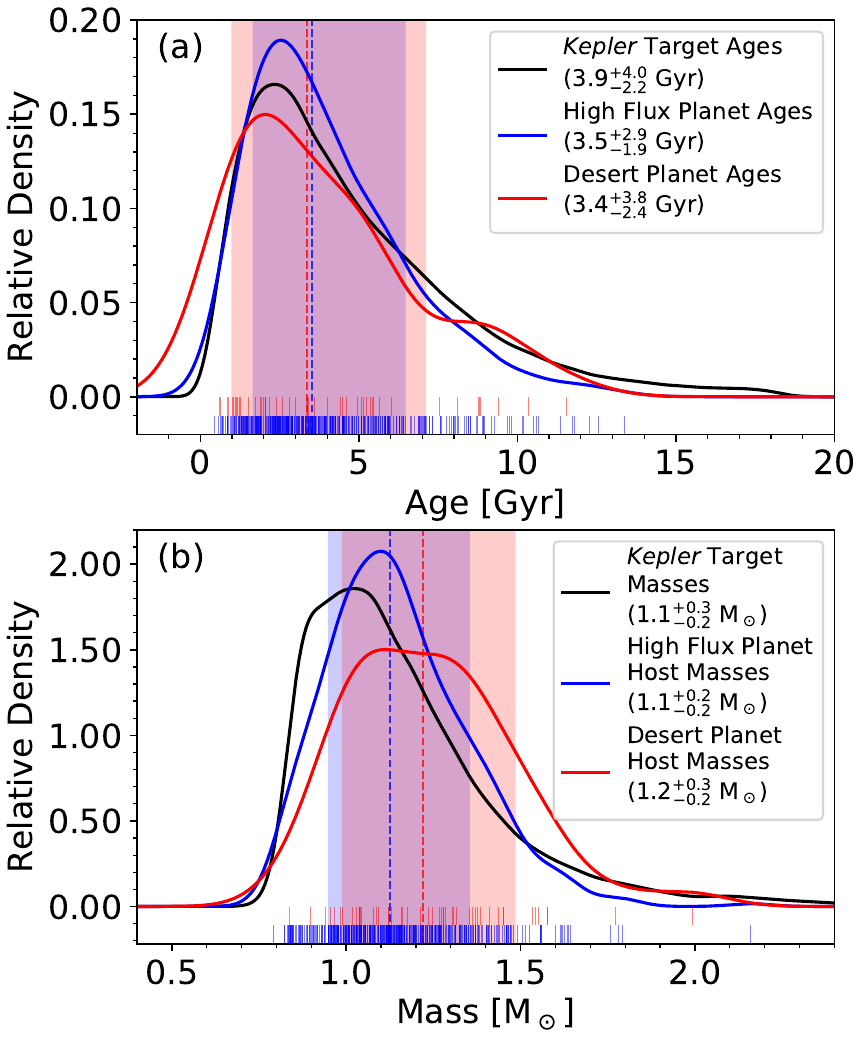}}
\caption{Stellar age (panel (a)) and mass (panel (b)) distributions for planets with high incident flux ($F_\mathrm{p}$\,$>$\,650\,\fearth) within the hot \replaced{super-Earth}{sub-Neptunian} desert \rp\,=\,2.2--3.8\,\rearth\ (red), at high incident flux outside the hot \replaced{super-Earth}{sub-Neptunian} desert (blue), and the overall \kep\ target sample with reliable ages (RUWE\,$<$\,1.2, \texttt{iso\_gof}\,$>$\,0.99, and TAMS\,$<$\,20\,Gyr, black). The blue and red ticks represent the individual ages/masses used to calculate the KDEs using bandwidths following Scott's Rule \citep{Scott1992}. The dashed vertical lines and shaded areas are the median and 1$\sigma$ bounds.}
\label{fig:desert}
\end{figure}

\begin{figure*}
\resizebox{0.8\textwidth}{!}{\includegraphics{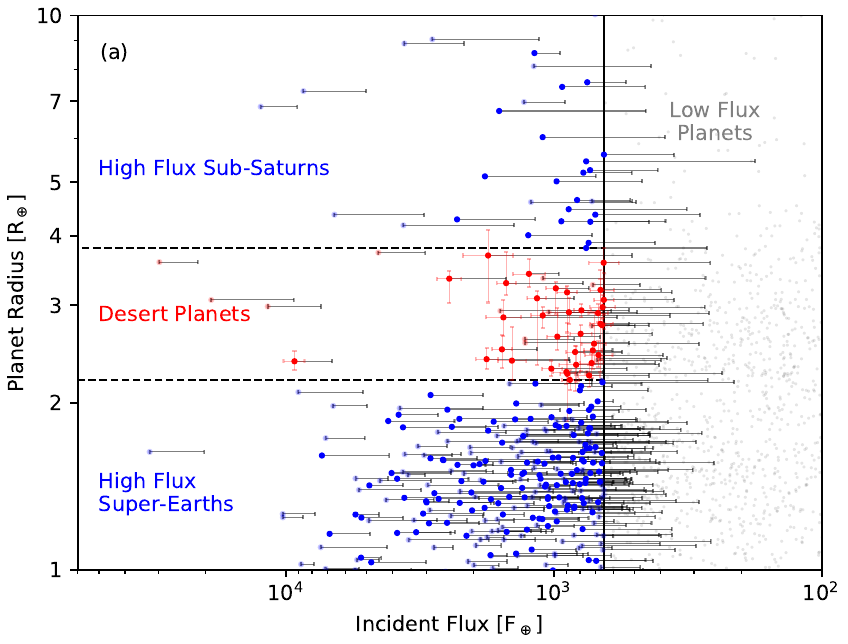}}
\resizebox{0.8\textwidth}{!}{\includegraphics{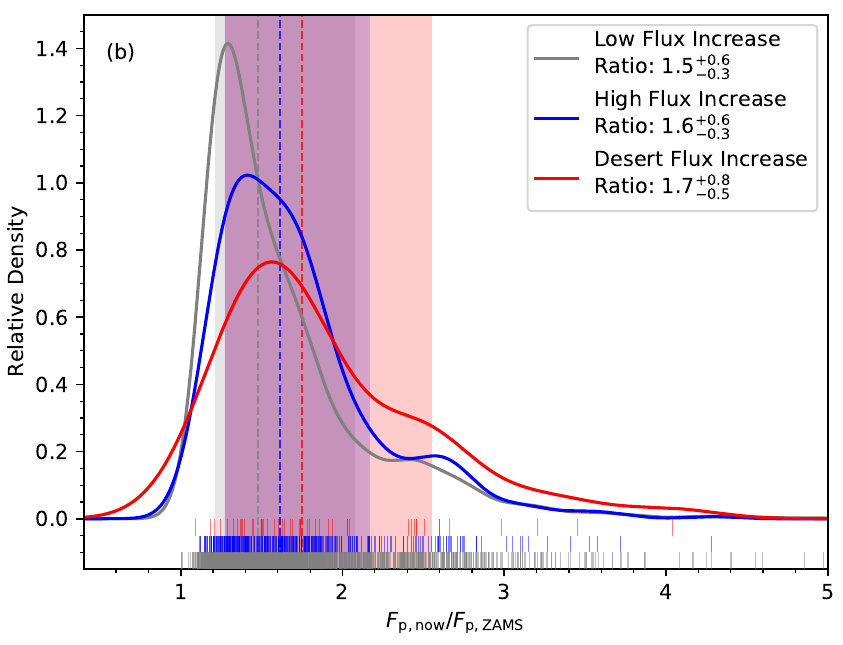}}
\centering
\caption{Panel (a): Planet radius versus incident flux for \kep\ exoplanets in the hot \replaced{super-Earth}{sub-Neptunian} desert (red) defined by \cite{Lundkvist2016}, at incident fluxes $>$\,650\,\fearth\ and outside the desert (blue), and at incident fluxes $<$\,650\,\fearth (grey). Solid and translucent points are confirmed and candidate planets, respectively. The black, butted bars show the incident flux history of each planet, starting at the incident flux the planet received at the zero age main sequence (ZAMS). Panel (b): Current divided by ZAMS flux ratio for desert planets (red), other high incident flux planets (blue), and all other planets (grey).}
\label{fig:desertevo}
\end{figure*}

The hot \replaced{super-Earth}{sub-Neptunian} desert, defined as $R_{\mathrm{p}}$\,=\,2.2--3.8\,\rearth\ and $F_{\mathrm{p}}$\,$>$\,650\,\fearth, is another region of parameter space believed to be devoid of planets \citep{Lundkvist2016}. The lack of planets in the hot \replaced{super-Earth}{sub-Neptunian} desert can be explained by photoevaporation \citep{Lopez2012,owen16,Owen2017}, with hydrogen and helium atmospheres being completely lost to the large EUV flux at small orbital separations \citep{Lopez2017}. At lower incident EUV fluxes, planets still lose significant portions of their atmospheres, but stabilize at an envelope mass fraction of 1--2\% \citep{Owen2017}. \cite{Berger2018c} found 74 confirmed planets/planet candidates in the ``desert'', and suggested that some of these planets (1) may be the remnants of the photoevaporation of a giant planet's envelope \citep{baraffe05}, (2) did not receive enough EUV flux to lose their low molecular weight atmospheres \citep{Owen2017}, or (3) may have high molecular weight atmospheres \citep{Lopez2017}. Our newly derived stellar masses and ages can shed additional light on the properties and formation of this intriguing class of planets.

Figure \ref{fig:desert}(a) shows the age distribution of planets within (red) and outside (blue) the hot \replaced{super-Earth}{sub-Neptunian} desert at high incident fluxes \citep[$>$\,650\,\fearth,][]{Lundkvist2016}. We also plot the overall \kep\ sample for comparison. We observe similar stellar age distributions for planets within the hot \replaced{super-Earth}{sub-Neptunian} desert and those at high incident fluxes with radii outside 2.2--3.8\,\rearth. The median ages and distributions are almost identical, which is also confirmed with a K-S test (p-value\,=\,0.14). We therefore conclude that most of these ``desert dwellers'' are not young planets that are currently losing mass.

Unlike stellar ages, Figure \ref{fig:desert}(b) indicates a difference between the stellar mass distributions of high incident flux planets inside and outside the desert. A K-S test yielded a p-value of 0.02, indicating a difference at $\approx$\,2$\sigma$ significance. Therefore, we tentatively conclude that desert planets tend to be around more massive stars. This conclusion is also supported by a K-S test using host star radii and uncertainties, yielding a p-value of 0.007. Because hot \replaced{super-Earth}{sub-Neptune} planet hosts appear to have higher stellar masses and larger stellar radii, we hypothesize that these planets could have shifted into the desert through stellar evolution.

To investigate this, Figure \ref{fig:desertevo}(a) shows the incident flux history of planets within and outside the desert. We count 35 confirmed and 15 planet candidates within the hot \replaced{super-Earth}{sub-Neptunian} desert \citep[as defined by][]{Lundkvist2016}, and we denote them with the ``Desert'' flag in Table \ref{tab:pars}. Figure \ref{fig:desertevo}(b) shows the ratios of the current flux compared to ZAMS flux (see Table \ref{tab:pars} for planets in the different regimes of panel (a). Together, Figures \ref{fig:desertevo}(a) and (b) suggest that 60\% of desert planets have moved into the desert as a result of stellar luminosity evolution. Therefore, we infer that the majority of desert planets were not, in the first $\sim$\,100\,Myr, exposed to enough EUV flux to completely strip their atmospheres. However, this incident flux evolution does not explain all desert planets.

We find nine confirmed planets within the desert by $>$\,1$\sigma$ even after accounting for the effects of stellar evolution on the incident flux:  \kep-234\,b, \kep-541\,b, \kep-611\,b, \kep-644\,b, \kep-645\,b, \kep-656\,b, \kep-1016\,b, \kep-1171\,b, and \kep-1518\,b. If we assume a ZAMS luminosity that corresponds to the ZAMS luminosity of the lower mass uncertainty bound for each system, we still find three planets within the desert:  \kep-644\,b, \kep-645\,b, and \kep-1171\,b. \kep-644 and \kep-645 have AO-detected companions, and thus their stellar properties are likely inaccurate, in addition to the incident fluxes their planets receive. Using ZAMS luminosities corresponding to the lower mass uncertainty bound is a pessimistic assumption, which overestimates the increase in the star's luminosity since the ZAMS and accounts for uncertainties. Given 100\,Myr H/He atmosphere-loss timescales \citep{Owen2017}, it is unlikely that these desert-dwelling planets are typical sub-Neptune-mass planets with H/He envelopes, unless they migrated to their current orbital separations \citep{dong18} and/or have higher-molecular weight atmospheres \citep{Lopez2017,Gaidos2020}. It is also possible that these planets are the bare cores of 2--3$\times$ more massive planets \citep{Armstrong2020}. While photoevaporation is not expected to strip enough atmospheric mass off of these massive cores, tidal disruption \citep{Vick2019} or giant planet collisions \citep{Mordasini2018} can produce the required mass loss. Alternatively, these bare cores may form in-situ by opening up a gas-less gap in the protoplanetary disk and avoiding runaway accretion \citep{Lee2019}. Ultimately, these desert-dwellers represent interesting tests of planet formation and evolution theories, and warrant additional scrutiny.

\section{Cool Planets}

\subsection{Cool, Inflated Jupiters} \label{sec:cooljup}

\begin{figure*}
\resizebox{\hsize}{!}{\includegraphics{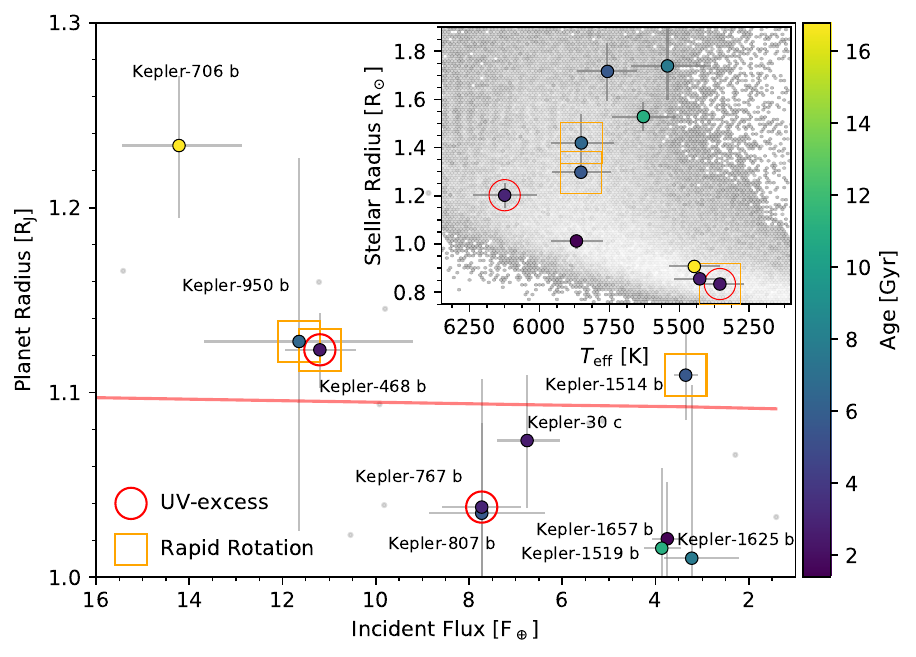}}
\caption{Radius versus incident flux for \kep's cool Jupiters. Planets are colored by isochrone age. Orange squares and red rings indicate stars that are rotating more rapidly or exhibit more UV-excess compared to the Hyades, respectively. Grey points are candidate planets. The red curve represents the maximum radius for a 4.5\,Gyr, Jupiter-mass, pure hydrogen and helium object \citep{Thorngren2016}. The inset shows the position of the host stars on the H-R diagram compared to the \kep\ target sample (grey).}
\label{fig:coolinfjup}
\end{figure*}

The mechanism producing the inflated radii of hot Jupiters is still a major unsolved problem \citep{Guillot2002,Fortney2010,fortney11,Baraffe2010,Baraffe2014,Laughlin2015,Laughlin2018,Komacek2020}. Most theories are linked to the observation that all inflated Jupiters experience high incident flux \citep{Demory2011,Laughlin2011,miller11}, whether they are orbiting main sequence stars or have been re-inflated from the effects of post-main sequence stellar evolution \citep{Lopez2016,Grunblatt2016,Grunblatt2017,Grunblatt2019}. Thus, finding examples of cool, inflated Jupiters may present an interesting challenge to these theories. \cite{Berger2018c} identified three Jupiters with anomalously large radii and at $<$\,150\,\fearth. If these planets are inflated at low incident fluxes, there must be some other mechanism causing their inflation. For example, these planets may be young and hot from the gravitational energy of accretion, and this additional energy could produce an inflated atmosphere \citep{Lopez2012}. These cool, inflated Jupiters could also be heated from recent tidal interactions with other planets and their host star \citep{Jackson2008,Fortney2010b}.

Similar to \cite{Berger2018c}, we find a small sample of cool, confirmed (or validated), $>$\,1\,$\mathrm{R_J}$ Jupiters (Figure \ref{fig:coolinfjup}). In addition to isochrone ages, we mark points according to whether they exhibit UV-excess (red circle) or rapid rotation (orange square) as additional indicators for youth. We flag a star as having UV-excess when it meets two criteria: (1) $m_{\mathrm{NUV}} - m_{K_s} < 8.3$ (to avoid magnitude-limited cases) and (2) $m_{\mathrm{NUV}} - m_{K_s} < m_{\mathrm{NUV}}(M_{K_s})_{\mathrm{Hyades}} - m_{K_s}(M_{K_s})_{\mathrm{Hyades}}$, where the condition is set by the Hyades relation evaluated at the $M_{K_s}$ of the \kep\ star. The Hyades cluster \citep[$\sim$\,650\,Myr][]{Boesgaard2016} has a well-defined trend in NUV--$K_s$ versus $M_{K_s}$, making it an effective separator for young/old stars in the \kep\ field \citep{Berger2018}. We obtained the NUV fluxes for the \kep\ and Hyades stars from the Galaxy Evolution Explorer \citep[GALEX;][]{GALEX,Olmedo2015}. Similarly, we flag a star as having rapid rotation if its rotation period is more rapid than the Hyades gyrochrone with an initial rotation period of 3.4\,days \citep{Kundert2012}. We use the rotation periods derived in \cite{McQuillan2013,McQuillan2014} or \cite{Mazeh2015}.

We find two planets that are significantly above the maximum radius for a 4.5\,Gyr, Jupiter-mass, pure hydrogen and helium object \citep{Thorngren2016}. \kep-468\,b appears to be young according to its host's isochrone age (2.4$^{+3.3}_{-1.7}$\,Gyr), excess UV flux \citep{Skumanich1972,Soderblom2010}, and rapid rotation period. It does not have a rotation period detection according to \cite{McQuillan2013,McQuillan2014} or \cite{Mazeh2015}, but an inspection of the \kep\ lightcurve reveals a rotation period of $\approx$\,5.7 days, consistent with a young age \citep{Barnes2010}. \kep-468 has RUWE\,=\,1.15, which is below the threshold for being a likely binary. Similarly, \kep-468 does not appear to have any companions according to the high resolution imaging \citep{law14,Furlan2017}. Even at 2.4\,Gyr, a Jupiter-mass planet may still be cooling and contracting from its heat from formation, although the majority of this contraction occurs within the first $\sim$\,Gyr \citep{Fortney2007,Linder2019}. Hence, it is possible that \kep-468\,b is young and still cooling and contracting from its heat from formation.

Unlike \kep-468\,b, \kep-706\,b does not appear to orbit a young star. \kep-706 has an isochrone age of $\sim$\,17\,Gyr, a rotation period of $\approx$\,38 days measured from the \kep\ lightcurve, and a NUV magnitude that is beyond the GALEX limiting magnitude of 22.6 \citep{Olmedo2015}. While the isochrone age is unreliable, the rotation period supports an old age for this star \citep{Barnes2010}. Furthermore, neither the RUWE value (1.13) nor high resolution imaging from \cite{law14} indicate the presence of a stellar companion.

If confirmed, the radius and age of \kep-706\,b would be highly interesting. Given its old age, it is very unlikely that it is inflated from its residual heat from formation based on the expected cooling timescales \citep{Lopez2012}. \kep-706\,b orbits every 41\,days, so it is also unlikely that its radius is inflated by strong tidal or magnetic interactions with its host star, barring a highly elliptical orbit. Planet-planet interactions typically produce only a fraction of the tidal heating that is caused by the host star \citep{Hay2019}, but this heating is strongly dependent on the proximity and the mass of the other planet. Rings are another potential explanation for \kep-706\,b's inflated radius \citep{Schlichting2011}, but we cannot confirm the presence of rings with \kep's lightcurve cadence/precision. Additional follow-up observations will be required to confirm whether \kep-706\,b is indeed a bona-fide cool, inflated Jupiter.

\subsection{Habitable Zone Planets} \label{sec:habzone}

Our work also yields a revised list of planets in the habitable zone. Following the definition of \cite{Kane2016}, we find 133 planet candidates and 32 confirmed planets within the habitable zone (0.25--1.5\,\fearth), all of which are flagged in Table \ref{tab:pars}. Compared to \cite{Berger2018c}, we count two fewer confirmed and 24 more candidate planets. We report eight confirmed planets with radii $<$\,2\,\rearth: \kep-62\,e, \kep-62\,f, \kep-186\,f, \kep-283\,c, \kep-440\,b, \kep-442\,b, \kep-452\,b \citep[but see also][]{Mullally18}, and \kep-1544\,b, although \kep-62, \kep-186, and \kep-283 have AO-detected companions \citep{Furlan2017}. This increase of habitable zone candidate planets is mostly caused by the slightly cooler \teff\ derived in \ref{B20}, which produce smaller incident fluxes for planets at the same orbital periods.

\section{Single and Multi Planet Systems} \label{sec:multis}

\begin{figure}
\resizebox{\hsize}{!}{\includegraphics{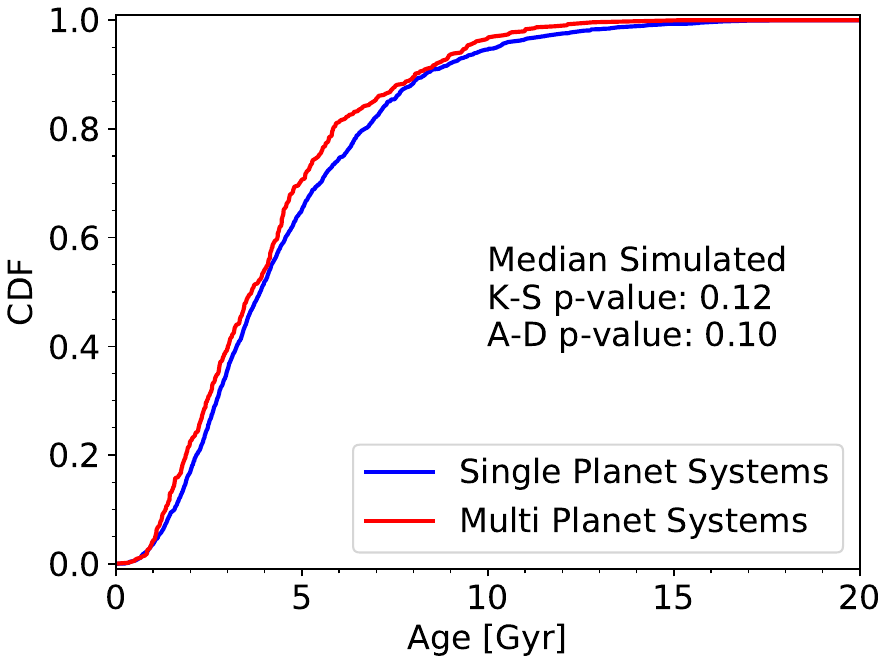}}
\caption{Cumulative age distribution functions (CDFs) of \kep\ systems with multiple transiting planets (red, 539 systems) and one transiting planet (blue, 1768 systems).}
\label{fig:multiage}
\end{figure}

Several observational results suggest that the \kep\ multiplanet systems (multis) have smaller eccentricities and inclinations than the \kep\ systems with just one small transiting planet (singles). The \kep\ multis are nearly coplanar \citep{Fang2012,Fabrycky2014}, with a typical mutual inclination of $1.0^\circ$-$2.2^\circ$. These typical mutual inclinations are smaller than what is required for all of the planets to transit, indicating that detection bias alone cannot explain the observed coplanarity of the \kep\ multis. Furthermore, the \kep\ singles have higher eccentricities than the multis, as determined from their more varied transit durations \citep{xie16,Mills2019,VanEylen2019}. In multis with an ultra-short period planet, the mutual inclinations between the planets are higher than in multis without an ultra-short period planet \citep{Dai2019}, suggesting that the evolutionary pathway that creates ultra-short period planets also excites planetary inclinations and eccentricities \citep[e.g.,][]{Petrovich2019}.

What physical processes affect the ecccentricities and inclinations of the \kep\ planets?  The host star masses, metallicities, and \vsini's of the \kep\ singles and multis are statistically indistinguishable \citep{Weiss2018b}, suggesting that the host stars are not the main source of the observed differences in the eccentricity and inclination distributions.  One possible source of dynamical excitation (and also instability) is the planets themselves: dynamically packed planetary systems become unstable on timescales of $10^{7}$-$10^{10}$ orbits \citep{Obertas2017}, with the closest-packed systems typically becoming unstable earliest.  Systems with larger eccentricities are more likely to become unstable, excited in inclination, and tidally circularized  \citep{Pu2019,Petrovich2019}.  Long-distance processes such as a gravitational perturbation from a passing star or can also affect the coplanarity and stability of the \kep\ multis \citep{Spalding2016}.  The likelihood that any of these disruptive mechanisms has already occurred increases with time.  Therefore, significant differences in stellar age between the single and multi-planet systems, in particular a preference for single planets to be around older stars, may point to an evolutionary pathway from the multis to the singles.

Figure \ref{fig:multiage} shows cumulative distribution functions for ages of multis (red, 539 systems) and singles (blue, 1768 systems). We observe that the single transiting planet systems indeed appear to be on average older than the multiple transiting planet systems. To quantify the difference of these two distributions, we conducted Monte Carlo K-S simulations to account for individual errors in the ages of our stars. We also performed Monte Carlo Anderson-Darling \citep[A-D, \texttt{scipy's anderson\_ksamp};][]{Scholz1987,Scipy} test simulations in a similar manner. The resulting p-values are 0.12 and 0.10 for the K-S and A-D tests, respectively, indicating that the single- and multi-planet ages are statistically indistinguishable. Requiring spectroscopic metallicities for single and multi systems, which improve isochrone ages \citep{Howes2019}, produced p-values of 0.59 and 0.25 (capped) for the K-S and A-D tests, respectively.

We therefore conclude that the ages of single and multiple transiting planet systems are statistically indistinguishable within the precision of our age measurements. This implies that dynamical interactions that affect planet eccentricities and inclinations such as planet-planet and planet-passing star interactions operate significantly faster than the $\sim$\,Gyr timescales that can be resolved by our sample \citep[e.g.,][]{Spalding2016,Obertas2017}, and may happen at almost any age. For instance, young singles could be the multis that were disrupted at young ages, while old multis have not been disrupted yet.

\section{Summary and Conclusions} \label{sec:conc}

We presented a re-analysis of \kep\ exoplanet properties using the uniform stellar properties provided by \ref{B20}. In particular, we performed a systematic analysis of planet properties as a function of stellar mass and age for the entire \kep\ sample. Our main conclusions are as follows:

\begin{itemize}

    \item We observe that the location of the planet radius gap increases in planet radius with increasing stellar mass, with a slope of \slope\ = $0.26^{+0.21}_{-0.16}$, consistent with previous results based on \replaced{a subsample by \cite{Fulton2018}}{smaller samples by \cite{Fulton2018} and \cite{Cloutier2020b}}. The uncertainty on the slope even for the full \kep\ sample is too large to discern between \replaced{photoevaporation}{a planet mass dependence on stellar mass \citep[as required by photoevaporation,][]{Wu2019}} and core-powered mass-loss \citep{Gupta2020}. We estimate that differentiating between these theories would require $\gtrsim$\,20,000 planets with 1\% fractional precisions in planet radius and stellar mass.  
    
    \item We find first evidence for a stellar age dependence of the fraction of super-Earths to sub-Neptunes, increasing from young planets ($<$\,1\,Gyr, 0.61\,$\pm$\,0.09) to old planets ($>$\,1\,Gyr, 1.00\,$\pm$\,0.10). This is consistent with predictions of the core-powered mass-loss mechanism that sub-Neptunes evolve to become super-Earths over Gyr timescales \citep{Gupta2020}, but we caution that our sample does not constrain evolution at times $\lesssim$\,100\,Myr due to photoevaporation \citep{Owen2017}. We also observe that the age dependence of the fraction of super-Earths to sub-Neptunes becomes stronger with higher incident flux. \added{Additionally, there is a marginally significant (p-value\,=\,0.11) decrease in the radii of cool ($F_{\mathrm{p}}$\,$<$\,150\,\fearth) sub-Neptunes, which may suggest that these sub-Neptunes have H/He envelopes as opposed to higher mean molecular weight atmospheres}. However, we caution that the latter results may be influenced by small number statistics and possible effects of undetected binary companions.
    
    \item We identify 35 confirmed planets and 15 planet candidates that occupy the hot \replaced{super-Earth}{sub-Neptunian} desert \citep{Lundkvist2016}. We determine that most of the desert planets orbit evolved stars, and thus it is unlikely that they are young planets currently undergoing mass loss. In addition, we identify nine planets that are within the desert even after accounting for stellar evolution: \kep-234\,b, \kep-541\,b, \kep-611\,b, \kep-644\,b, \kep-645\,b, \kep-656\,b, \kep-1016\,b, \kep-1171\,b, and \kep-1518\,b.
    
    \item We investigate \kep's population of cool ($F_{\mathrm{p}}$ $<$\,20\,\fearth) Jupiters, and identify \kep-706\,b and \kep-468\,b as candidates for cool, inflated Jupiters. While \kep-468\,b orbits a young star and is potentially inflated due to its heat from formation, \kep-706\,b apparently orbits an old star based on its rotation period. Future observations will be required to rule out binary companions that could bias the radius measurement.
    
    \item We find that 32 confirmed and 133 planet candidates have incident fluxes from 0.25--1.50\,\fearth\ and thus occupy a nominal habitable zone \citep{Kane2016}. Of these, 37 planet candidates and eight confirmed planets have radii smaller than 2\,\rearth \citep[but see also][]{Mullally18,Burke2019}:  \kep-62\,e, \kep-62\,f, \kep-186\,f, \kep-283\,c, \kep-440\,b, \kep-442\,b, \kep-452\,b, and \kep-1544\,b.
    
    \item We find that stars hosting multiple transiting planets have ages that are on average higher but are statistically indistinguishable from those of single planet host stars. This implies that if dynamical interactions (planet-planet and/or planet-passing star interactions) frequently scatter planets out of our line-of-sight to produce single transiting planet systems, these interactions are quick \citep[e.g.,][]{Spalding2016,Obertas2017} and may happen over a wide range of ages.
    
\end{itemize}

Our results demonstrate the importance of precise, homogeneous parameters and the power of stellar ages and masses to allow a more comprehensive investigation of exoplanet populations and their evolution over time. An extension to lower mass stars will require the use of alternative age indicators such as lithium abundances, rotation, and UV-excess measurements to identify a more robust sample of young \kep\ hosts. Additionally, a homogeneous stellar classification of planet hosts observed by $K2$ and $TESS$ will offer new insights into their planet populations and provide additional clues about planet formation and evolution.

\begin{deluxetable*}{lrcrrrrr}
\tabletypesize{\scriptsize}
\tablenum{1}
\tablewidth{0pt}
\tablecolumns{8}
\tablecaption{Planet Parameters}
\tablehead{
\colhead{KIC ID} & \colhead{KOI ID} & \colhead{Planet Disposition} & \colhead{Planet Radius [\rearth]} & \colhead{Semimajor Axis [AU]} & \colhead{Incident Flux [\fearth]} & \colhead{ZAMS Flux [\fearth]} & \colhead{Interesting Object Flag}}
\def\arraystretch{1.0}
\startdata
11446443&1.01&CONFIRMED&14.21$^{+0.29}_{-0.29}$&0.0355$^{+0.0008}_{-0.0008}$&854.78$^{+69.34}_{-64.81}$&524.75&AO\\
10666592&2.01&CONFIRMED&16.45$^{+0.35}_{-0.34}$&0.0381$^{+0.0005}_{-0.0006}$&4285.95$^{+331.23}_{-336.99}$&2394.68&AO\\
10748390&3.01&CONFIRMED&4.88$^{+0.08}_{-0.07}$&0.0516$^{+0.0006}_{-0.0004}$&86.58$^{+5.24}_{-4.73}$&58.22&\\
3861595&4.01&CONFIRMED&13.18$^{+0.42}_{-0.82}$&0.0583$^{+0.0007}_{-0.0015}$&5244.42$^{+478.90}_{-520.46}$&3647.63&AO\\
11853905&7.01&CONFIRMED&4.01$^{+0.10}_{-0.10}$&0.0455$^{+0.0007}_{-0.0010}$&1247.17$^{+99.09}_{-99.02}$&719.80&\\
6922244&10.01&CONFIRMED&15.53$^{+0.36}_{-0.36}$&0.0490$^{+0.0007}_{-0.0008}$&1482.86$^{+117.24}_{-126.09}$&965.36&AO\\
5812701&12.01&CONFIRMED&14.36$^{+0.32}_{-0.29}$&0.1509$^{+0.0018}_{-0.0021}$&204.64$^{+18.91}_{-15.30}$&141.77&YoungAO\\
10874614&17.01&CONFIRMED&12.76$^{+0.27}_{-0.32}$&0.0445$^{+0.0007}_{-0.0009}$&793.33$^{+56.16}_{-61.47}$&450.14&\\
8191672&18.01&CONFIRMED&14.91$^{+0.43}_{-0.42}$&0.0504$^{+0.0008}_{-0.0007}$&1826.02$^{+162.50}_{-151.51}$&1079.02&AO\\
11804465&20.01&CONFIRMED&19.46$^{+0.45}_{-0.42}$&0.0553$^{+0.0012}_{-0.0014}$&837.81$^{+68.80}_{-69.12}$&425.39&\\
9631995&22.01&CONFIRMED&12.83$^{+0.26}_{-0.26}$&0.0818$^{+0.0014}_{-0.0017}$&277.28$^{+20.78}_{-20.87}$&181.79&\\
6521045&41.01&CONFIRMED&2.34$^{+0.27}_{-0.09}$&0.1120$^{+0.0026}_{-0.0029}$&192.74$^{+14.67}_{-14.70}$&91.63&AO\\
6521045&41.02&CONFIRMED&1.35$^{+0.03}_{-0.11}$&0.0740$^{+0.0017}_{-0.0019}$&441.16$^{+33.57}_{-33.64}$&209.73&AO\\
6521045&41.03&CONFIRMED&1.60$^{+0.33}_{-0.09}$&0.2202$^{+0.0050}_{-0.0056}$&49.86$^{+3.79}_{-3.80}$&23.70&AO\\
8866102&42.01&CONFIRMED&2.76$^{+0.06}_{-0.05}$&0.1427$^{+0.0021}_{-0.0029}$&147.65$^{+10.00}_{-9.35}$&91.23&AO\\
10905239&46.01&CONFIRMED&6.03$^{+0.17}_{-0.16}$&0.0468$^{+0.0007}_{-0.0006}$&1102.03$^{+89.48}_{-79.55}$&385.33&\\
\enddata
\tablecomments{The interesting object flag denotes whether the host star is included in the older (``Old'') or younger (``Young'') than 1\,Gyr samples selected in \S\ref{sec:pradage}, whether the planets are located within the valley (``Gap''; see \S\ref{sec:ingap}), whether the planet is located within the hot \replaced{super-Earth}{sub-Neptunian} desert (``Desert''; see \S\ref{sec:desert}), whether planets are in the habitable zone (``HZ''; see \S\ref{sec:habzone}), and/or whether the host has an AO-detected companion \citep{Furlan2017}. A subset of our planet parameters is provided here to illustrate the form and format. The full table, in machine-readable format, can be found online.} \label{tab:pars}
\end{deluxetable*}

\acknowledgments 
We gratefully acknowledge everyone involved in the \gaia\ and \kep\ missions for their tireless efforts which have made this paper possible. \added{We thank Yanqin Wu, James Owen, and Edwin Kite for helpful comments that improved the paper.} We thank Erik Petigura, Sam Grunblatt, Jamie Tayar, Ashley Chontos, Erica Bufanda, Maryum Sayeed, Connor Auge, Vanshree Bhalotia, Nicholas Saunders, Michael Liu, Benjamin Boe, and Ehsan Kourkchi for helpful discussions in addition to feedback on the figures. T.A.B. and D.H. acknowledge support by a NASA FINESST award (80NSSC19K1424) and the National Science Foundation (AST-1717000). D.H. also acknowledges support from the Alfred P. Sloan Foundation. E.G. acknowledges support from NSF award AST-187215 and, as a visiting professor to the University of Göttingen, the German Science Foundation through DFG Research 644 Unit FOR2544 “Blue Planets around Red Stars". This research was partially conducted during the Exostar19 program at the Kavli Institute for Theoretical Physics at UC Santa Barbara, which was supported in part by the National Science Foundation under Grant No. NSF PHY-1748958. This work has made use of data from the European Space Agency (ESA) mission {\it Gaia} (\url{https://www.cosmos.esa.int/gaia}), processed by the {\it Gaia} Data Processing and Analysis Consortium (DPAC, \url{https://www.cosmos.esa.int/web/gaia/dpac/consortium}). Funding for the DPAC has been provided by national institutions, in particular the institutions participating in the {\it Gaia} Multilateral Agreement. This publication makes use of data products from the Two Micron All Sky Survey, which is a joint project of the University of Massachusetts and the Infrared Processing and Analysis Center/California Institute of Technology, funded by the National Aeronautics and Space Administration and the National Science Foundation. This research has made use of NASA's Astrophysics Data System. This research was made possible through the use of the AAVSO Photometric All-Sky Survey (APASS), funded by the Robert Martin Ayers Sciences Fund. This research made use of the cross-match service provided by CDS, Strasbourg. This research has made use of the NASA Exoplanet Archive, which is operated by the California Institute of Technology, under contract with the National Aeronautics and Space Administration under the Exoplanet Exploration Program.

\vspace{5mm}

\software{\texttt{astropy} \citep{astropy},
		  \texttt{dustmaps} \citep{Green2018,Green2019},
		  \texttt{gapfit} \citep{Loyd2020},
		  \texttt{GNU Parallel} \citep{Tange2018},
		  \texttt{isoclassify} \citep{huber17},
		  \texttt{KS2D} \citep{Gabinou2018},
		  \texttt{Matplotlib} \citep{Matplotlib},
          \texttt{mwdust} \citep{bovy16}, 
          \texttt{Pandas} \citep{Pandas}, 
          \texttt{scikit-learn} \citep{scikit-learn},
          \texttt{SciPy} \citep{Scipy},
          \texttt{skimage} \citep{skimage}}

\bibliography{references}

\newcommand{\SortNoop}[1]{}
\begin{thebibliography}{}
\expandafter\ifx\csname natexlab\endcsname\relax\def\natexlab#1{#1}\fi
\providecommand{\url}[1]{\href{#1}{#1}}

\bibitem[{{Armitage} \& {Rice}(2005)}]{Armitage2005}
{Armitage}, P.~J., \& {Rice}, W.~K.~M. 2005, arXiv e-prints, astro

\bibitem[{{Armstrong} {et~al.}(2020){Armstrong}, {Lopez}, {Adibekyan}, {Booth},
  {Bryant}, {Collins}, {Emsenhuber}, {Huang}, {King}, {Lillo-box}, {Lissauer},
  {Matthews}, {Mousis}, {Nielsen}, {Osborn}, {Otegi}, {Santos}, {Sousa},
  {Stassun}, {Veras}, {Ziegler}, {Acton}, {Almenara}, {Anderson}, {Barrado},
  {Barros}, {Bayliss}, {Belardi}, {Bouchy}, {Briceno}, {Brogi}, {Brown},
  {Burleigh}, {Casewell}, {Chaushev}, {Ciardi}, {Collins}, {Col{\'o}n},
  {Cooke}, {Crossfield}, {D{\'\i}az}, {Deleuil}, {Delgado Mena}, {Demangeon},
  {Dorn}, {Dumusque}, {Eigmuller}, {Fausnaugh}, {Figueira}, {Gan}, {Gand hi},
  {Gill}, {Goad}, {Guenther}, {Helled}, {Hojjatpanah}, {Howell}, {Jackman},
  {Jenkins}, {Jenkins}, {Jensen}, {Kennedy}, {Latham}, {Law}, {Lendl},
  {Lozovsky}, {Mann}, {Moyano}, {McCormac}, {Meru}, {Mordasini}, {Osborn},
  {Pollacco}, {Queloz}, {Raynard}, {Ricker}, {Rowden}, {Santerne}, {Schlieder},
  {Seager}, {Sha}, {Tan}, {Tilbrook}, {Ting}, {Udry}, {Vanderspek}, {Watson},
  {West}, {Wilson}, {Winn}, {Wheatley}, {Villasenor}, {Vines}, \&
  {Zhan}}]{Armstrong2020}
{Armstrong}, D.~J., {Lopez}, T.~A., {Adibekyan}, V., {et~al.} 2020, arXiv
  e-prints, arXiv:2003.10314

\bibitem[{{Astropy Collaboration} {et~al.}(2013){Astropy Collaboration},
  {Robitaille}, {Tollerud}, {Greenfield}, {Droettboom}, {Bray}, {Aldcroft},
  {Davis}, {Ginsburg}, {Price-Whelan}, {Kerzendorf}, {Conley}, {Crighton},
  {Barbary}, {Muna}, {Ferguson}, {Grollier}, {Parikh}, {Nair}, {Unther},
  {Deil}, {Woillez}, {Conseil}, {Kramer}, {Turner}, {Singer}, {Fox}, {Weaver},
  {Zabalza}, {Edwards}, {Azalee Bostroem}, {Burke}, {Casey}, {Crawford},
  {Dencheva}, {Ely}, {Jenness}, {Labrie}, {Lim}, {Pierfederici}, {Pontzen},
  {Ptak}, {Refsdal}, {Servillat}, \& {Streicher}}]{astropy}
{Astropy Collaboration}, {Robitaille}, T.~P., {Tollerud}, E.~J., {et~al.} 2013,
  \aap, 558, A33

\bibitem[{{Baraffe} {et~al.}(2010){Baraffe}, {Chabrier}, \&
  {Barman}}]{Baraffe2010}
{Baraffe}, I., {Chabrier}, G., \& {Barman}, T. 2010, Reports on Progress in
  Physics, 73, 016901

\bibitem[{{Baraffe} {et~al.}(2005){Baraffe}, {Chabrier}, {Barman}, {Selsis},
  {Allard}, \& {Hauschildt}}]{baraffe05}
{Baraffe}, I., {Chabrier}, G., {Barman}, T.~S., {et~al.} 2005, \aap, 436, L47

\bibitem[{{Baraffe} {et~al.}(2014){Baraffe}, {Chabrier}, {Fortney}, \&
  {Sotin}}]{Baraffe2014}
{Baraffe}, I., {Chabrier}, G., {Fortney}, J., \& {Sotin}, C. 2014, in
  Protostars and Planets VI, ed. H.~{Beuther}, R.~S. {Klessen}, C.~P.
  {Dullemond}, \& T.~{Henning}, 763

\bibitem[{{Barnes} \& {Kim}(2010)}]{Barnes2010}
{Barnes}, S.~A., \& {Kim}, Y.-C. 2010, \apj, 721, 675

\bibitem[{{Batalha} {et~al.}(2010){Batalha}, {Borucki}, {Koch}, {Bryson},
  {Haas}, {Brown}, {Caldwell}, {Hall}, {Gilliland}, {Latham}, {Meibom}, \&
  {Monet}}]{batalha10}
{Batalha}, N.~M., {Borucki}, W.~J., {Koch}, D.~G., {et~al.} 2010, \apjl, 713,
  L109

\bibitem[{{Berger} {et~al.}(2018{\natexlab{a}}){Berger}, {Howard}, \&
  {Boesgaard}}]{Berger2018}
{Berger}, T.~A., {Howard}, A.~W., \& {Boesgaard}, A.~M. 2018{\natexlab{a}},
  \apj, 855, 115

\bibitem[{{Berger} {et~al.}(2018{\natexlab{b}}){Berger}, {Huber}, {Gaidos}, \&
  {van Saders}}]{Berger2018c}
{Berger}, T.~A., {Huber}, D., {Gaidos}, E., \& {van Saders}, J.~L.
  2018{\natexlab{b}}, \apj, 866, 99

\bibitem[{{Berger} {et~al.}(2020){Berger}, {Huber}, {van Saders}, {Gaidos},
  {Tayar}, \& {Kraus}}]{Berger2020a}
{Berger}, T.~A., {Huber}, D., {van Saders}, J.~L., {et~al.} 2020, \aj, 159, 280

\bibitem[{{Boesgaard} {et~al.}(2016){Boesgaard}, {Lum}, {Deliyannis}, {King},
  {Pinsonneault}, \& {Somers}}]{Boesgaard2016}
{Boesgaard}, A.~M., {Lum}, M.~G., {Deliyannis}, C.~P., {et~al.} 2016, \apj,
  830, 49

\bibitem[{{Bovy} {et~al.}(2016){Bovy}, {Rix}, {Green}, {Schlafly}, \&
  {Finkbeiner}}]{bovy16}
{Bovy}, J., {Rix}, H.-W., {Green}, G.~M., {Schlafly}, E.~F., \& {Finkbeiner},
  D.~P. 2016, \apj, 818, 130

\bibitem[{{Bryson} {et~al.}(2020){Bryson}, {Coughlin}, {Batalha}, {Berger},
  {Huber}, {Burke}, {Dotson}, \& {Mullally}}]{Bryson2020}
{Bryson}, S., {Coughlin}, J., {Batalha}, N.~M., {et~al.} 2020, \aj, 159, 279

\bibitem[{{Burke} {et~al.}(2019){Burke}, {Mullally}, {Thompson}, {Coughlin}, \&
  {Rowe}}]{Burke2019}
{Burke}, C.~J., {Mullally}, F., {Thompson}, S.~E., {Coughlin}, J.~L., \&
  {Rowe}, J.~F. 2019, \aj, 157, 143

\bibitem[{{Choi} {et~al.}(2016){Choi}, {Dotter}, {Conroy}, {Cantiello},
  {Paxton}, \& {Johnson}}]{choi16}
{Choi}, J., {Dotter}, A., {Conroy}, C., {et~al.} 2016, \apj, 823, 102

\bibitem[{{Christiansen} {et~al.}(2012){Christiansen}, {Jenkins}, {Caldwell},
  {Burke}, {Tenenbaum}, {Seader}, {Thompson}, {Barclay}, {Clarke}, {Li},
  {Smith}, {Stumpe}, {Twicken}, \& {Van Cleve}}]{Christiansen2012}
{Christiansen}, J.~L., {Jenkins}, J.~M., {Caldwell}, D.~A., {et~al.} 2012,
  \pasp, 124, 1279

\bibitem[{{Claret} \& {Bloemen}(2011)}]{claret11}
{Claret}, A., \& {Bloemen}, S. 2011, \aap, 529, A75

\bibitem[{{Cloutier} \& {Menou}(2020)}]{Cloutier2020b}
{Cloutier}, R., \& {Menou}, K. 2020, \aj, 159, 211

\bibitem[{{Cloutier} {et~al.}(2020){Cloutier}, {Eastman}, {Rodriguez},
  {Astudillo-Defru}, {Bonfils}, {Mortier}, {Watson}, {Stalport}, {Pinamonti},
  {Lienhard}, {Harutyunyan}, {Damasso}, {Latham}, {Collins}, {Massey}, {Irwin},
  {Winters}, {Charbonneau}, {Ziegler}, {Matthews}, {Crossfield}, {Kreidberg},
  {Quinn}, {Ricker}, {Vanderspek}, {Seager}, {Winn}, {Jenkins}, {Vezie},
  {Udry}, {Twicken}, {Tenenbaum}, {Sozzetti}, {S{\'e}gransan}, {Schlieder},
  {Sasselov}, {Santos}, {Rice}, {Rackham}, {Poretti}, {Piotto}, {Phillips},
  {Pepe}, {Molinari}, {Mignon}, {Micela}, {Melo}, {de Medeiros}, {Mayor},
  {Matson}, {Martinez Fiorenzano}, {Mann}, {Magazz{\'u}}, {Lovis},
  {L{\'o}pez-Morales}, {Lopez}, {Lissauer}, {L{\'e}pine}, {Law}, {Kielkopf},
  {Johnson}, {Jensen}, {Howell}, {Gonzales}, {Ghedina}, {Forveille},
  {Figueira}, {Dumusque}, {Dressing}, {Doyon}, {D{\'\i}az}, {Fabrizio},
  {Delfosse}, {Cosentino}, {Conti}, {Collins}, {Cameron}, {Ciardi}, {Caldwell},
  {Burke}, {Buchhave}, {Brice{\~n}o}, {Boyd}, {Bouchy}, {Beichman}, {Artigau},
  \& {Almenara}}]{Cloutier2020}
{Cloutier}, R., {Eastman}, J.~D., {Rodriguez}, J.~E., {et~al.} 2020, \aj, 160,
  3

\bibitem[{{Dai} {et~al.}(2019){Dai}, {Masuda}, {Winn}, \& {Zeng}}]{Dai2019}
{Dai}, F., {Masuda}, K., {Winn}, J.~N., \& {Zeng}, L. 2019, \apj, 883, 79

\bibitem[{{Demory} \& {Seager}(2011)}]{Demory2011}
{Demory}, B.-O., \& {Seager}, S. 2011, \apjs, 197, 12

\bibitem[{{Dong} {et~al.}(2018){Dong}, {Xie}, {Zhou}, {Zheng}, \&
  {Luo}}]{dong18}
{Dong}, S., {Xie}, J.-W., {Zhou}, J.-L., {Zheng}, Z., \& {Luo}, A. 2018,
  Proceedings of the National Academy of Science, 115, 266

\bibitem[{{Dotter}(2016)}]{dotter16}
{Dotter}, A. 2016, \apjs, 222, 8

\bibitem[{{Eker} {et~al.}(2018){Eker}, {Bak{\i}{\c s}}, {Bilir}, {Soydugan},
  {Steer}, {Soydugan}, {Bak{\i}{\c s}}, {Ali{\c c}avu{\c s}}, {Aslan}, \&
  {Alpsoy}}]{Eker2018}
{Eker}, Z., {Bak{\i}{\c s}}, V., {Bilir}, S., {et~al.} 2018, \mnras, 479, 5491

\bibitem[{{Epstein} \& {Pinsonneault}(2014)}]{Epstein2014}
{Epstein}, C.~R., \& {Pinsonneault}, M.~H. 2014, \apj, 780, 159

\bibitem[{{Evans}(2018)}]{Evans2018}
{Evans}, D.~F. 2018, Research Notes of the American Astronomical Society, 2, 20

\bibitem[{{Everett} {et~al.}(2013){Everett}, {Howell}, {Silva}, \&
  {Szkody}}]{everett13}
{Everett}, M.~E., {Howell}, S.~B., {Silva}, D.~R., \& {Szkody}, P. 2013, \apj,
  771, 107

\bibitem[{{Fabrycky} {et~al.}(2014){Fabrycky}, {Lissauer}, {Ragozzine}, {Rowe},
  {Steffen}, {Agol}, {Barclay}, {Batalha}, {Borucki}, {Ciardi}, {Ford},
  {Gautier}, {Geary}, {Holman}, {Jenkins}, {Li}, {Morehead}, {Morris},
  {Shporer}, {Smith}, {Still}, \& {Van Cleve}}]{Fabrycky2014}
{Fabrycky}, D.~C., {Lissauer}, J.~J., {Ragozzine}, D., {et~al.} 2014, \apj,
  790, 146

\bibitem[{{Fang} \& {Margot}(2012)}]{Fang2012}
{Fang}, J., \& {Margot}, J.-L. 2012, \apj, 761, 92

\bibitem[{{Fortney} {et~al.}(2010){Fortney}, {Baraffe}, \&
  {Militzer}}]{Fortney2010b}
{Fortney}, J.~J., {Baraffe}, I., \& {Militzer}, B. 2010, {Giant Planet Interior
  Structure and Thermal Evolution}, ed. S.~{Seager}, 397--418

\bibitem[{{Fortney} {et~al.}(2007){Fortney}, {Marley}, \&
  {Barnes}}]{Fortney2007}
{Fortney}, J.~J., {Marley}, M.~S., \& {Barnes}, J.~W. 2007, \apj, 659, 1661

\bibitem[{{Fortney} \& {Nettelmann}(2010)}]{Fortney2010}
{Fortney}, J.~J., \& {Nettelmann}, N. 2010, \ssr, 152, 423

\bibitem[{{Fortney} {et~al.}(2011){Fortney}, {Demory}, {D{\'e}sert}, {Rowe},
  {Marcy}, {Isaacson}, {Buchhave}, {Ciardi}, {Gautier}, {Batalha}, {Caldwell},
  {Bryson}, {Nutzman}, {Jenkins}, {Howard}, {Charbonneau}, {Knutson}, {Howell},
  {Everett}, {Fressin}, {Deming}, {Borucki}, {Brown}, {Ford}, {Gilliland},
  {Latham}, {Miller}, {Seager}, {Fischer}, {Koch}, {Lissauer}, {Haas}, {Still},
  {Lucas}, {Gillon}, {Christiansen}, \& {Geary}}]{fortney11}
{Fortney}, J.~J., {Demory}, B.-O., {D{\'e}sert}, J.-M., {et~al.} 2011, \apjs,
  197, 9

\bibitem[{{Fulton} \& {Petigura}(2018)}]{Fulton2018}
{Fulton}, B.~J., \& {Petigura}, E.~A. 2018, \aj, 156, 264

\bibitem[{{Fulton} {et~al.}(2017){Fulton}, {Petigura}, {Howard}, {Isaacson},
  {Marcy}, {Cargile}, {Hebb}, {Weiss}, {Johnson}, {Morton}, {Sinukoff},
  {Crossfield}, \& {Hirsch}}]{Fulton2017}
{Fulton}, B.~J., {Petigura}, E.~A., {Howard}, A.~W., {et~al.} 2017, \aj, 154,
  109

\bibitem[{{Furlan} {et~al.}(2017){Furlan}, {Ciardi}, {Everett}, {Saylors},
  {Teske}, {Horch}, {Howell}, {van Belle}, {Hirsch}, {Gautier}, {Adams},
  {Barrado}, {Cartier}, {Dressing}, {Dupree}, {Gilliland}, {Lillo-Box},
  {Lucas}, \& {Wang}}]{Furlan2017}
{Furlan}, E., {Ciardi}, D.~R., {Everett}, M.~E., {et~al.} 2017, \aj, 153, 71

\bibitem[{{Gaia Collaboration} {et~al.}(2018){Gaia Collaboration}, {Brown},
  {Vallenari}, {Prusti}, {de Bruijne}, {Babusiaux}, {Bailer-Jones}, {Biermann},
  {Evans}, {Eyer}, {Jansen}, {Jordi}, {Klioner}, {Lammers}, {Lindegren},
  {Luri}, {Mignard}, {Panem}, {Pourbaix}, {Randich}, {Sartoretti}, {Siddiqui},
  {Soubiran}, {van Leeuwen}, {Walton}, {Arenou}, {Bastian}, {Cropper},
  {Drimmel}, {Katz}, {Lattanzi}, {Bakker}, {Cacciari}, {Casta{\~n}eda},
  {Chaoul}, {Cheek}, {De Angeli}, {Fabricius}, {Guerra}, {Holl}, {Masana},
  {Messineo}, {Mowlavi}, {Nienartowicz}, {Panuzzo}, {Portell}, {Riello},
  {Seabroke}, {Tanga}, {Th{\'e}venin}, {Gracia-Abril}, {Comoretto},
  {Garcia-Reinaldos}, {Teyssier}, {Altmann}, {Andrae}, {Audard},
  {Bellas-Velidis}, {Benson}, {Berthier}, {Blomme}, {Burgess}, {Busso},
  {Carry}, {Cellino}, {Clementini}, {Clotet}, {Creevey}, {Davidson}, {De
  Ridder}, {Delchambre}, {Dell'Oro}, {Ducourant},
  {Fern{\'a}ndez-Hern{\'a}ndez}, {Fouesneau}, {Fr{\'e}mat}, {Galluccio},
  {Garc{\'\i}a-Torres}, {Gonz{\'a}lez-N{\'u}{\~n}ez}, {Gonz{\'a}lez-Vidal},
  {Gosset}, {Guy}, {Halbwachs}, {Hambly}, {Harrison}, {Hern{\'a}ndez},
  {Hestroffer}, {Hodgkin}, {Hutton}, {Jasniewicz}, {Jean-Antoine-Piccolo},
  {Jordan}, {Korn}, {Krone-Martins}, {Lanzafame}, {Lebzelter}, {L{\"o}ffler},
  {Manteiga}, {Marrese}, {Mart{\'\i}n-Fleitas}, {Moitinho}, {Mora}, {Muinonen},
  {Osinde}, {Pancino}, {Pauwels}, {Petit}, {Recio-Blanco}, {Richards},
  {Rimoldini}, {Robin}, {Sarro}, {Siopis}, {Smith}, {Sozzetti}, {S{\"u}veges},
  {Torra}, {van Reeven}, {Abbas}, {Abreu Aramburu}, {Accart}, {Aerts},
  {Altavilla}, {{\'A}lvarez}, {Alvarez}, {Alves}, {Anderson}, {Andrei},
  {Anglada Varela}, {Antiche}, {Antoja}, {Arcay}, {Astraatmadja}, {Bach},
  {Baker}, {Balaguer-N{\'u}{\~n}ez}, {Balm}, {Barache}, {Barata}, {Barbato},
  {Barblan}, {Barklem}, {Barrado}, {Barros}, {Barstow}, {Bartholom{\'e}
  Mu{\~n}oz}, {Bassilana}, {Becciani}, {Bellazzini}, {Berihuete}, {Bertone},
  {Bianchi}, {Bienaym{\'e}}, {Blanco-Cuaresma}, {Boch}, {Boeche}, {Bombrun},
  {Borrachero}, {Bossini}, {Bouquillon}, {Bourda}, {Bragaglia}, {Bramante},
  {Breddels}, {Bressan}, {Brouillet}, {Br{\"u}semeister}, {Brugaletta},
  {Bucciarelli}, {Burlacu}, {Busonero}, {Butkevich}, {Buzzi}, {Caffau},
  {Cancelliere}, {Cannizzaro}, {Cantat-Gaudin}, {Carballo}, {Carlucci},
  {Carrasco}, {Casamiquela}, {Castellani}, {Castro-Ginard}, {Charlot},
  {Chemin}, {Chiavassa}, {Cocozza}, {Costigan}, {Cowell}, {Crifo}, {Crosta},
  {Crowley}, {Cuypers}, {Dafonte}, {Damerdji}, {Dapergolas}, {David}, {David},
  {de Laverny}, {De Luise}, {De March}, {de Martino}, {de Souza}, {de Torres},
  {Debosscher}, {del Pozo}, {Delbo}, {Delgado}, {Delgado}, {Di Matteo},
  {Diakite}, {Diener}, {Distefano}, {Dolding}, {Drazinos}, {Dur{\'a}n},
  {Edvardsson}, {Enke}, {Eriksson}, {Esquej}, {Eynard Bontemps}, {Fabre},
  {Fabrizio}, {Faigler}, {Falc{\~a}o}, {Farr{\`a}s Casas}, {Federici},
  {Fedorets}, {Fernique}, {Figueras}, {Filippi}, {Findeisen}, {Fonti},
  {Fraile}, {Fraser}, {Fr{\'e}zouls}, {Gai}, {Galleti}, {Garabato},
  {Garc{\'\i}a-Sedano}, {Garofalo}, {Garralda}, {Gavel}, {Gavras}, {Gerssen},
  {Geyer}, {Giacobbe}, {Gilmore}, {Girona}, {Giuffrida}, {Glass}, {Gomes},
  {Granvik}, {Gueguen}, {Guerrier}, {Guiraud}, {Guti{\'e}rrez-S{\'a}nchez},
  {Haigron}, {Hatzidimitriou}, {Hauser}, {Haywood}, {Heiter}, {Helmi}, {Heu},
  {Hilger}, {Hobbs}, {Hofmann}, {Holland}, {Huckle}, {Hypki}, {Icardi},
  {Jan{\ss}en}, {Jevardat de Fombelle}, {Jonker}, {Juh{\'a}sz}, {Julbe},
  {Karampelas}, {Kewley}, {Klar}, {Kochoska}, {Kohley}, {Kolenberg},
  {Kontizas}, {Kontizas}, {Koposov}, {Kordopatis}, {Kostrzewa-Rutkowska},
  {Koubsky}, {Lambert}, {Lanza}, {Lasne}, {Lavigne}, {Le Fustec}, {Le
  Poncin-Lafitte}, {Lebreton}, {Leccia}, {Leclerc}, {Lecoeur-Taibi},
  {Lenhardt}, {Leroux}, {Liao}, {Licata}, {Lindstr{\o}m}, {Lister}, {Livanou},
  {Lobel}, {L{\'o}pez}, {Managau}, {Mann}, {Mantelet}, {Marchal}, {Marchant},
  {Marconi}, {Marinoni}, {Marschalk{\'o}}, {Marshall}, {Martino}, {Marton},
  {Mary}, {Massari}, {Matijevi{\v{c}}}, {Mazeh}, {McMillan}, {Messina},
  {Michalik}, {Millar}, {Molina}, {Molinaro}, {Moln{\'a}r}, {Montegriffo},
  {Mor}, {Morbidelli}, {Morel}, {Morris}, {Mulone}, {Muraveva}, {Musella},
  {Nelemans}, {Nicastro}, {Noval}, {O'Mullane}, {Ord{\'e}novic},
  {Ord{\'o}{\~n}ez-Blanco}, {Osborne}, {Pagani}, {Pagano}, {Pailler},
  {Palacin}, {Palaversa}, {Panahi}, {Pawlak}, {Piersimoni}, {Pineau}, {Plachy},
  {Plum}, {Poggio}, {Poujoulet}, {Pr{\v{s}}a}, {Pulone}, {Racero}, {Ragaini},
  {Rambaux}, {Ramos-Lerate}, {Regibo}, {Reyl{\'e}}, {Riclet}, {Ripepi}, {Riva},
  {Rivard}, {Rixon}, {Roegiers}, {Roelens}, {Romero-G{\'o}mez}, {Rowell},
  {Royer}, {Ruiz-Dern}, {Sadowski}, {Sagrist{\`a} Sell{\'e}s}, {Sahlmann},
  {Salgado}, {Salguero}, {Sanna}, {Santana-Ros}, {Sarasso}, {Savietto},
  {Schultheis}, {Sciacca}, {Segol}, {Segovia}, {S{\'e}gransan}, {Shih},
  {Siltala}, {Silva}, {Smart}, {Smith}, {Solano}, {Solitro}, {Sordo}, {Soria
  Nieto}, {Souchay}, {Spagna}, {Spoto}, {Stampa}, {Steele},
  {Steidelm{\"u}ller}, {Stephenson}, {Stoev}, {Suess}, {Surdej}, {Szabados},
  {Szegedi-Elek}, {Tapiador}, {Taris}, {Tauran}, {Taylor}, {Teixeira},
  {Terrett}, {Teyssand ier}, {Thuillot}, {Titarenko}, {Torra Clotet}, {Turon},
  {Ulla}, {Utrilla}, {Uzzi}, {Vaillant}, {Valentini}, {Valette}, {van Elteren},
  {Van Hemelryck}, {van Leeuwen}, {Vaschetto}, {Vecchiato}, {Veljanoski},
  {Viala}, {Vicente}, {Vogt}, {von Essen}, {Voss}, {Votruba}, {Voutsinas},
  {Walmsley}, {Weiler}, {Wertz}, {Wevers}, {Wyrzykowski}, {Yoldas},
  {{\v{Z}}erjal}, {Ziaeepour}, {Zorec}, {Zschocke}, {Zucker}, {Zurbach}, \&
  {Zwitter}}]{Brown2018}
{Gaia Collaboration}, {Brown}, A.~G.~A., {Vallenari}, A., {et~al.} 2018, \aap,
  616, A1

\bibitem[{{Gaidos} \& {Mann}(2013)}]{gaidos13}
{Gaidos}, E., \& {Mann}, A.~W. 2013, \apj, 762, 41

\bibitem[{{Gaidos} {et~al.}(2020){Gaidos}, {Hirano}, {Mann}, {Owens}, {Berger},
  {France}, {Vanderburg}, {Harakawa}, {Hodapp}, {Ishizuka}, {Jacobson},
  {Konishi}, {Kotani}, {Kudo}, {Kurokawa}, {Kuzuhara}, {Nishikawa}, {Omiya},
  {Serizawa}, {Tamura}, \& {Ueda}}]{Gaidos2020}
{Gaidos}, E., {Hirano}, T., {Mann}, A.~W., {et~al.} 2020, \mnras, 495, 650

\bibitem[{{Ginzburg} {et~al.}(2016){Ginzburg}, {Schlichting}, \&
  {Sari}}]{Ginzburg2016}
{Ginzburg}, S., {Schlichting}, H.~E., \& {Sari}, R. 2016, \apj, 825, 29

\bibitem[{{Ginzburg} {et~al.}(2018){Ginzburg}, {Schlichting}, \&
  {Sari}}]{Ginzburg2018}
---. 2018, \mnras, 476, 759

\bibitem[{{Green} {et~al.}(2019){Green}, {Schlafly}, {Zucker}, {Speagle}, \&
  {Finkbeiner}}]{Green2019}
{Green}, G.~M., {Schlafly}, E., {Zucker}, C., {Speagle}, J.~S., \&
  {Finkbeiner}, D. 2019, \apj, 887, 93

\bibitem[{{Green} {et~al.}(2018){Green}, {Schlafly}, {Finkbeiner}, {Rix},
  {Martin}, {Burgett}, {Draper}, {Flewelling}, {Hodapp}, {Kaiser}, {Kudritzki},
  {Magnier}, {Metcalfe}, {Tonry}, {Wainscoat}, \& {Waters}}]{Green2018}
{Green}, G.~M., {Schlafly}, E.~F., {Finkbeiner}, D., {et~al.} 2018, \mnras,
  478, 651

\bibitem[{{Grunblatt} {et~al.}(2019){Grunblatt}, {Huber}, {Gaidos}, {Hon},
  {Zinn}, \& {Stello}}]{Grunblatt2019}
{Grunblatt}, S.~K., {Huber}, D., {Gaidos}, E., {et~al.} 2019, \aj, 158, 227

\bibitem[{{Grunblatt} {et~al.}(2016){Grunblatt}, {Huber}, {Gaidos}, {Lopez},
  {Fulton}, {Vanderburg}, {Barclay}, {Fortney}, {Howard}, {Isaacson}, {Mann},
  {Petigura}, {Silva Aguirre}, \& {Sinukoff}}]{Grunblatt2016}
{Grunblatt}, S.~K., {Huber}, D., {Gaidos}, E.~J., {et~al.} 2016, \aj, 152, 185

\bibitem[{{Grunblatt} {et~al.}(2017){Grunblatt}, {Huber}, {Gaidos}, {Lopez},
  {Howard}, {Isaacson}, {Sinukoff}, {Vanderburg}, {Nofi}, {Yu}, {North},
  {Chaplin}, {Foreman-Mackey}, {Petigura}, {Ansdell}, {Weiss}, {Fulton}, \&
  {Lin}}]{Grunblatt2017}
{Grunblatt}, S.~K., {Huber}, D., {Gaidos}, E., {et~al.} 2017, \aj, 154, 254

\bibitem[{{Guillot} \& {Showman}(2002)}]{Guillot2002}
{Guillot}, T., \& {Showman}, A.~P. 2002, \aap, 385, 156

\bibitem[{{Gupta} \& {Schlichting}(2019)}]{Gupta2019}
{Gupta}, A., \& {Schlichting}, H.~E. 2019, \mnras, 487, 24

\bibitem[{{Gupta} \& {Schlichting}(2020)}]{Gupta2020}
---. 2020, \mnras, 493, 792

\bibitem[{{Hardegree-Ullman} {et~al.}(2020){Hardegree-Ullman}, {Zink},
  {Christiansen}, {Dressing}, {Ciardi}, \& {Schlieder}}]{Hardegree2020}
{Hardegree-Ullman}, K.~K., {Zink}, J.~K., {Christiansen}, J.~L., {et~al.} 2020,
  \apjs, 247, 28

\bibitem[{{Hay} \& {Matsuyama}(2019)}]{Hay2019}
{Hay}, H. C.~F.~C., \& {Matsuyama}, I. 2019, \apj, 875, 22

\bibitem[{{Howe} \& {Burrows}(2015)}]{Howe2015}
{Howe}, A.~R., \& {Burrows}, A. 2015, \apj, 808, 150

\bibitem[{{Howell} {et~al.}(2014){Howell}, {Sobeck}, {Haas}, {Still},
  {Barclay}, {Mullally}, {Troeltzsch}, {Aigrain}, {Bryson}, {Caldwell},
  {Chaplin}, {Cochran}, {Huber}, {Marcy}, {Miglio}, {Najita}, {Smith},
  {Twicken}, \& {Fortney}}]{Howell2014}
{Howell}, S.~B., {Sobeck}, C., {Haas}, M., {et~al.} 2014, \pasp, 126, 398

\bibitem[{{Howes} {et~al.}(2019){Howes}, {Lindegren}, {Feltzing}, {Church}, \&
  {Bensby}}]{Howes2019}
{Howes}, L.~M., {Lindegren}, L., {Feltzing}, S., {Church}, R.~P., \& {Bensby},
  T. 2019, \aap, 622, A27

\bibitem[{{Huber} {et~al.}(2014){Huber}, {Silva Aguirre}, {Matthews},
  {Pinsonneault}, {Gaidos}, {Garc{\'{\i}}a}, {Hekker}, {Mathur}, {Mosser},
  {Torres}, {Bastien}, {Basu}, {Bedding}, {Chaplin}, {Demory}, {Fleming},
  {Guo}, {Mann}, {Rowe}, {Serenelli}, {Smith}, \& {Stello}}]{huber14}
{Huber}, D., {Silva Aguirre}, V., {Matthews}, J.~M., {et~al.} 2014, \apjs, 211,
  2

\bibitem[{{Huber} {et~al.}(2017){Huber}, {Zinn}, {Bojsen-Hansen},
  {Pinsonneault}, {Sahlholdt}, {Serenelli}, {Silva Aguirre}, {Stassun},
  {Stello}, {Tayar}, {Bastien}, {Bedding}, {Buchhave}, {Chaplin}, {Davies},
  {Garc{\'{\i}}a}, {Latham}, {Mathur}, {Mosser}, \& {Sharma}}]{huber17}
{Huber}, D., {Zinn}, J., {Bojsen-Hansen}, M., {et~al.} 2017, \apj, 844, 102

\bibitem[{Hunter(2007)}]{Matplotlib}
Hunter, J.~D. 2007, Computing In Science \& Engineering, 9, 90

\bibitem[{{Jackson} {et~al.}(2008){Jackson}, {Greenberg}, \&
  {Barnes}}]{Jackson2008}
{Jackson}, B., {Greenberg}, R., \& {Barnes}, R. 2008, \apj, 681, 1631

\bibitem[{{Johnson} {et~al.}(2017){Johnson}, {Petigura}, {Fulton}, {Marcy},
  {Howard}, {Isaacson}, {Hebb}, {Cargile}, {Morton}, {Weiss}, {Winn}, {Rogers},
  {Sinukoff}, \& {Hirsch}}]{Johnson2017}
{Johnson}, J.~A., {Petigura}, E.~A., {Fulton}, B.~J., {et~al.} 2017, \aj, 154,
  108

\bibitem[{{Kane} {et~al.}(2016){Kane}, {Hill}, {Kasting}, {Kopparapu},
  {Quintana}, {Barclay}, {Batalha}, {Borucki}, {Ciardi}, {Haghighipour},
  {Hinkel}, {Kaltenegger}, {Selsis}, \& {Torres}}]{Kane2016}
{Kane}, S.~R., {Hill}, M.~L., {Kasting}, J.~F., {et~al.} 2016, \apj, 830, 1

\bibitem[{{Kite} {et~al.}(2020){Kite}, {Fegley}, {Schaefer}, \&
  {Ford}}]{Kite2020}
{Kite}, E.~S., {Fegley}, Bruce, J., {Schaefer}, L., \& {Ford}, E.~B. 2020,
  \apj, 891, 111

\bibitem[{{Komacek} {et~al.}(2020){Komacek}, {Thorngren}, {Lopez}, \&
  {Ginzburg}}]{Komacek2020}
{Komacek}, T.~D., {Thorngren}, D.~P., {Lopez}, E.~D., \& {Ginzburg}, S. 2020,
  \apj, 893, 36

\bibitem[{{Kundert} {et~al.}(2012){Kundert}, {Cargile}, {Dhital}, {Hebb},
  {Rostron}, \& {Stassun}}]{Kundert2012}
{Kundert}, A., {Cargile}, P.~A., {Dhital}, S., {et~al.} 2012, in American
  Astronomical Society Meeting Abstracts, Vol. 219, American Astronomical
  Society Meeting Abstracts \#219, 152.13

\bibitem[{{Laughlin}(2018)}]{Laughlin2018}
{Laughlin}, G. 2018, {Mass-Radius Relations of Giant Planets: The Radius
  Anomaly and Interior Models}, 1

\bibitem[{{Laughlin} {et~al.}(2011){Laughlin}, {Crismani}, \&
  {Adams}}]{Laughlin2011}
{Laughlin}, G., {Crismani}, M., \& {Adams}, F.~C. 2011, \apjl, 729, L7

\bibitem[{{Laughlin} \& {Lissauer}(2015)}]{Laughlin2015}
{Laughlin}, G., \& {Lissauer}, J.~J. 2015, arXiv e-prints, arXiv:1501.05685

\bibitem[{{Law} {et~al.}(2014){Law}, {Morton}, {Baranec}, {Riddle},
  {Ravichandran}, {Ziegler}, {Johnson}, {Tendulkar}, {Bui}, {Burse}, {Das},
  {Dekany}, {Kulkarni}, {Punnadi}, \& {Ramaprakash}}]{law14}
{Law}, N.~M., {Morton}, T., {Baranec}, C., {et~al.} 2014, \apj, 791, 35

\bibitem[{{Lee}(2019)}]{Lee2019}
{Lee}, E.~J. 2019, \apj, 878, 36

\bibitem[{{Lee} \& {Chiang}(2016)}]{Lee2016}
{Lee}, E.~J., \& {Chiang}, E. 2016, \apj, 817, 90

\bibitem[{{Lee} {et~al.}(2014){Lee}, {Chiang}, \& {Ormel}}]{Lee2014}
{Lee}, E.~J., {Chiang}, E., \& {Ormel}, C.~W. 2014, \apj, 797, 95

\bibitem[{{Lindegren} {et~al.}(2018){Lindegren}, {Hern{\'a}ndez}, {Bombrun},
  {Klioner}, {Bastian}, {Ramos-Lerate}, {de Torres}, {Steidelm{\"u}ller},
  {Stephenson}, {Hobbs}, {Lammers}, {Biermann}, {Geyer}, {Hilger}, {Michalik},
  {Stampa}, {McMillan}, {Casta{\~n}eda}, {Clotet}, {Comoretto}, {Davidson},
  {Fabricius}, {Gracia}, {Hambly}, {Hutton}, {Mora}, {Portell}, {van Leeuwen},
  {Abbas}, {Abreu}, {Altmann}, {Andrei}, {Anglada}, {Balaguer-N{\'u}{\~n}ez},
  {Barache}, {Becciani}, {Bertone}, {Bianchi}, {Bouquillon}, {Bourda},
  {Br{\"u}semeister}, {Bucciarelli}, {Busonero}, {Buzzi}, {Cancelliere},
  {Carlucci}, {Charlot}, {Cheek}, {Crosta}, {Crowley}, {de Bruijne}, {de
  Felice}, {Drimmel}, {Esquej}, {Fienga}, {Fraile}, {Gai}, {Garralda},
  {Gonz{\'a}lez-Vidal}, {Guerra}, {Hauser}, {Hofmann}, {Holl}, {Jordan},
  {Lattanzi}, {Lenhardt}, {Liao}, {Licata}, {Lister}, {L{\"o}ffler},
  {Marchant}, {Martin-Fleitas}, {Messineo}, {Mignard}, {Morbidelli}, {Poggio},
  {Riva}, {Rowell}, {Salguero}, {Sarasso}, {Sciacca}, {Siddiqui}, {Smart},
  {Spagna}, {Steele}, {Taris}, {Torra}, {van Elteren}, {van Reeven}, \&
  {Vecchiato}}]{Lindegren2018}
{Lindegren}, L., {Hern{\'a}ndez}, J., {Bombrun}, A., {et~al.} 2018, \aap, 616,
  A2

\bibitem[{{Linder} {et~al.}(2019){Linder}, {Mordasini}, {Molli{\`e}re},
  {Marleau}, {Malik}, {Quanz}, \& {Meyer}}]{Linder2019}
{Linder}, E.~F., {Mordasini}, C., {Molli{\`e}re}, P., {et~al.} 2019, \aap, 623,
  A85

\bibitem[{{Lopez}(2017)}]{Lopez2017}
{Lopez}, E.~D. 2017, \mnras, 472, 245

\bibitem[{{Lopez} \& {Fortney}(2014)}]{Lopez2014}
{Lopez}, E.~D., \& {Fortney}, J.~J. 2014, \apj, 792, 1

\bibitem[{{Lopez} \& {Fortney}(2016)}]{Lopez2016}
---. 2016, \apj, 818, 4

\bibitem[{{Lopez} {et~al.}(2012){Lopez}, {Fortney}, \& {Miller}}]{Lopez2012}
{Lopez}, E.~D., {Fortney}, J.~J., \& {Miller}, N. 2012, \apj, 761, 59

\bibitem[{{Lopez} \& {Rice}(2018)}]{Lopez2018}
{Lopez}, E.~D., \& {Rice}, K. 2018, \mnras, 479, 5303

\bibitem[{{Loyd} {et~al.}(2020){Loyd}, {Shkolnik}, {Schneider},
  {Richey-Yowell}, {Barman}, {Peacock}, \& {Pagano}}]{Loyd2020}
{Loyd}, R.~O.~P., {Shkolnik}, E.~L., {Schneider}, A.~C., {et~al.} 2020, \apj,
  890, 23

\bibitem[{{Lundkvist} {et~al.}(2016){Lundkvist}, {Kjeldsen}, {Albrecht},
  {Davies}, {Basu}, {Huber}, {Justesen}, {Karoff}, {Silva Aguirre}, {van
  Eylen}, {Vang}, {Arentoft}, {Barclay}, {Bedding}, {Campante}, {Chaplin},
  {Christensen-Dalsgaard}, {Elsworth}, {Gilliland}, {Handberg}, {Hekker},
  {Kawaler}, {Lund}, {Metcalfe}, {Miglio}, {Rowe}, {Stello}, {Tingley}, \&
  {White}}]{Lundkvist2016}
{Lundkvist}, M.~S., {Kjeldsen}, H., {Albrecht}, S., {et~al.} 2016, Nature
  Communications, 7, 11201

\bibitem[{{Mandel} \& {Agol}(2002)}]{mandel02}
{Mandel}, K., \& {Agol}, E. 2002, \apjl, 580, L171

\bibitem[{{Mann} {et~al.}(2018){Mann}, {Vanderburg}, {Rizzuto}, {Kraus},
  {Berlind}, {Bieryla}, {Calkins}, {Esquerdo}, {Latham}, {Mace}, {Morris},
  {Quinn}, {Sokal}, \& {Stefanik}}]{Mann2018}
{Mann}, A.~W., {Vanderburg}, A., {Rizzuto}, A.~C., {et~al.} 2018, \aj, 155, 4

\bibitem[{{Martin} {et~al.}(2005){Martin}, {Fanson}, {Schiminovich},
  {Morrissey}, {Friedman}, {Barlow}, {Conrow}, {Grange}, {Jelinsky},
  {Milliard}, {Siegmund}, {Bianchi}, {Byun}, {Donas}, {Forster}, {Heckman},
  {Lee}, {Madore}, {Malina}, {Neff}, {Rich}, {Small}, {Surber}, {Szalay},
  {Welsh}, \& {Wyder}}]{GALEX}
{Martin}, D.~C., {Fanson}, J., {Schiminovich}, D., {et~al.} 2005, \apjl, 619,
  L1

\bibitem[{{Martinez} {et~al.}(2019){Martinez}, {Cunha}, {Ghezzi}, \&
  {Smith}}]{Martinez2019}
{Martinez}, C.~F., {Cunha}, K., {Ghezzi}, L., \& {Smith}, V.~V. 2019, \apj,
  875, 29

\bibitem[{{Mathur} {et~al.}(2017){Mathur}, {Huber}, {Batalha}, {Ciardi},
  {Bastien}, {Bieryla}, {Buchhave}, {Cochran}, {Endl}, {Esquerdo}, {Furlan},
  {Howard}, {Howell}, {Isaacson}, {Latham}, {MacQueen}, \&
  {Silva}}]{Mathur2017}
{Mathur}, S., {Huber}, D., {Batalha}, N.~M., {et~al.} 2017, \apjs, 229, 30

\bibitem[{{Mazeh} {et~al.}(2015){Mazeh}, {Perets}, {McQuillan}, \&
  {Goldstein}}]{Mazeh2015}
{Mazeh}, T., {Perets}, H.~B., {McQuillan}, A., \& {Goldstein}, E.~S. 2015,
  \apj, 801, 3

\bibitem[{McKinney(2010)}]{Pandas}
McKinney, W. 2010, in Proceedings of the 9th Python in Science Conference, ed.
  S.~van~der Walt \& J.~Millman, 51 -- 56

\bibitem[{{McQuillan} {et~al.}(2013){McQuillan}, {Mazeh}, \&
  {Aigrain}}]{McQuillan2013}
{McQuillan}, A., {Mazeh}, T., \& {Aigrain}, S. 2013, \apjl, 775, L11

\bibitem[{{McQuillan} {et~al.}(2014){McQuillan}, {Mazeh}, \&
  {Aigrain}}]{McQuillan2014}
---. 2014, \apjs, 211, 24

\bibitem[{{Miller} \& {Fortney}(2011)}]{miller11}
{Miller}, N., \& {Fortney}, J.~J. 2011, \apjl, 736, L29

\bibitem[{{Mills} {et~al.}(2019){Mills}, {Howard}, {Weiss}, {Steffen},
  {Isaacson}, {Fulton}, {Petigura}, {Kosiarek}, {Hirsch}, \&
  {Boisvert}}]{Mills2019}
{Mills}, S.~M., {Howard}, A.~W., {Weiss}, L.~M., {et~al.} 2019, \aj, 157, 145

\bibitem[{{Moe} \& {Di Stefano}(2017)}]{Moe2017}
{Moe}, M., \& {Di Stefano}, R. 2017, \apjs, 230, 15

\bibitem[{{Mordasini}(2018)}]{Mordasini2018}
{Mordasini}, C. 2018, {Planetary Population Synthesis}, 143

\bibitem[{{Mullally} {et~al.}(2018){Mullally}, {Thompson}, {Coughlin}, {Burke},
  \& {Rowe}}]{Mullally18}
{Mullally}, F., {Thompson}, S.~E., {Coughlin}, J.~L., {Burke}, C.~J., \&
  {Rowe}, J.~F. 2018, \aj, 155, 210

\bibitem[{{Nettelmann} {et~al.}(2011){Nettelmann}, {Fortney}, {Kramm}, \&
  {Redmer}}]{Nettelmann2011}
{Nettelmann}, N., {Fortney}, J.~J., {Kramm}, U., \& {Redmer}, R. 2011, \apj,
  733, 2

\bibitem[{{Nowak} {et~al.}(2020){Nowak}, {Luque}, {Parviainen}, {Pall{\'e}},
  {Molaverdikhani}, {B{\'e}jar}, {Lillo-Box}, {Rodr{\'\i}guez-L{\'o}pez},
  {Caballero}, {Zechmeister}, {Passegger}, {Cifuentes}, {Schweitzer}, {Narita},
  {Cale}, {Espinoza}, {Murgas}, {Zapatero Osorio}, {Pozuelos}, {Aceituno},
  {Amado}, {Barkaoui}, {Barrado}, {Bauer}, {Benkhaldoun}, {Caldwell},
  {Casasayas Barris}, {Chaturvedi}, {Chen}, {Collins}, {Collins},
  {Cort{\'e}s-Contreras}, {Crossfield}, {de Le{\'o}n}, {D{\'\i}ez Alonso},
  {Dreizler}, {El Mufti}, {Esparza-Borges}, {Essack}, {Fukui}, {Gillon},
  {Guerra}, {Hatzes}, {Henning}, {Herrero}, {Hesse}, {Hirano}, {Howell},
  {Jeffers}, {Jehin}, {Jenkins}, {Kaminski}, {Kemmer}, {Kielkopf},
  {Kossakowski}, {Kotani}, {K{\"u}rster}, {Lafarga}, {Latham}, {Law},
  {Lissauer}, {Lodieu}, {Madrigal-Aguado}, {Mann}, {Massey}, {Matson},
  {Matthews}, {Monta{\~n}{\'e}s-Rodr{\'\i}guez}, {Montes}, {Morales}, {Mori},
  {Nagel}, {Oshagh}, {Pedraz}, {Plavchan}, {Pollacco}, {Quirrenbach},
  {Reffert}, {Reiners}, {Ribas}, {Rose}, {Schlecker}, {Schlieder}, {Seager},
  {Stangret}, {Stock}, {Tamura}, {Teske}, {Trifonov}, {Twicken}, {Watanabe},
  {Wittrock}, {Ziegler}, \& {Zohrabi}}]{Nowak2020}
{Nowak}, G., {Luque}, R., {Parviainen}, H., {et~al.} 2020, arXiv e-prints,
  arXiv:2003.01140

\bibitem[{{Obertas} {et~al.}(2017){Obertas}, {Van Laerhoven}, \&
  {Tamayo}}]{Obertas2017}
{Obertas}, A., {Van Laerhoven}, C., \& {Tamayo}, D. 2017, \icarus, 293, 52

\bibitem[{{Olmedo} {et~al.}(2015){Olmedo}, {Lloyd}, {Mamajek}, {Ch{\'a}vez},
  {Bertone}, {Martin}, \& {Neill}}]{Olmedo2015}
{Olmedo}, M., {Lloyd}, J., {Mamajek}, E.~E., {et~al.} 2015, \apj, 813, 100

\bibitem[{{Owen}(2019)}]{Owen2019}
{Owen}, J.~E. 2019, Annual Review of Earth and Planetary Sciences, 47, 67

\bibitem[{{Owen} \& {Murray-Clay}(2018)}]{Owen2018}
{Owen}, J.~E., \& {Murray-Clay}, R. 2018, \mnras, 480, 2206

\bibitem[{{Owen} \& {Wu}(2013)}]{owen13}
{Owen}, J.~E., \& {Wu}, Y. 2013, \apj, 775, 105

\bibitem[{{Owen} \& {Wu}(2016)}]{owen16}
---. 2016, \apj, 817, 107

\bibitem[{{Owen} \& {Wu}(2017)}]{Owen2017}
---. 2017, \apj, 847, 29

\bibitem[{{Paxton} {et~al.}(2011){Paxton}, {Bildsten}, {Dotter}, {Herwig},
  {Lesaffre}, \& {Timmes}}]{paxton11}
{Paxton}, B., {Bildsten}, L., {Dotter}, A., {et~al.} 2011, \apjs, 192, 3

\bibitem[{{Paxton} {et~al.}(2013){Paxton}, {Cantiello}, {Arras}, {Bildsten},
  {Brown}, {Dotter}, {Mankovich}, {Montgomery}, {Stello}, {Timmes}, \&
  {Townsend}}]{paxton13}
{Paxton}, B., {Cantiello}, M., {Arras}, P., {et~al.} 2013, \apjs, 208, 4

\bibitem[{{Paxton} {et~al.}(2015){Paxton}, {Marchant}, {Schwab}, {Bauer},
  {Bildsten}, {Cantiello}, {Dessart}, {Farmer}, {Hu}, {Langer}, {Townsend},
  {Townsley}, \& {Timmes}}]{paxton15}
{Paxton}, B., {Marchant}, P., {Schwab}, J., {et~al.} 2015, \apjs, 220, 15

\bibitem[{Pedregosa {et~al.}(2011)Pedregosa, Varoquaux, Gramfort, Michel,
  Thirion, Grisel, Blondel, Prettenhofer, Weiss, Dubourg, Vanderplas, Passos,
  Cournapeau, Brucher, Perrot, \& Duchesnay}]{scikit-learn}
Pedregosa, F., Varoquaux, G., Gramfort, A., {et~al.} 2011, Journal of Machine
  Learning Research, 12, 2825

\bibitem[{{Petigura} {et~al.}(2017){Petigura}, {Howard}, {Marcy}, {Johnson},
  {Isaacson}, {Cargile}, {Hebb}, {Fulton}, {Weiss}, {Morton}, {Winn}, {Rogers},
  {Sinukoff}, {Hirsch}, \& {Crossfield}}]{Petigura2017}
{Petigura}, E.~A., {Howard}, A.~W., {Marcy}, G.~W., {et~al.} 2017, \aj, 154,
  107

\bibitem[{{Petigura} {et~al.}(2018){Petigura}, {Marcy}, {Winn}, {Weiss},
  {Fulton}, {Howard}, {Sinukoff}, {Isaacson}, {Morton}, \&
  {Johnson}}]{Petigura2018}
{Petigura}, E.~A., {Marcy}, G.~W., {Winn}, J.~N., {et~al.} 2018, \aj, 155, 89

\bibitem[{{Petrovich} {et~al.}(2019){Petrovich}, {Deibert}, \&
  {Wu}}]{Petrovich2019}
{Petrovich}, C., {Deibert}, E., \& {Wu}, Y. 2019, \aj, 157, 180

\bibitem[{{Pinsonneault} {et~al.}(2018){Pinsonneault}, {Elsworth}, {Tayar},
  {Serenelli}, {Stello}, {Zinn}, {Mathur}, {Garc{\'\i}a}, {Johnson}, {Hekker},
  {Huber}, {Kallinger}, {M{\'e}sz{\'a}ros}, {Mosser}, {Stassun}, {Girardi},
  {Rodrigues}, {Silva Aguirre}, {An}, {Basu}, {Chaplin}, {Corsaro}, {Cunha},
  {Garc{\'\i}a-Hern{\'a}ndez}, {Holtzman}, {J{\"o}nsson}, {Shetrone}, {Smith},
  {Sobeck}, {Stringfellow}, {Zamora}, {Beers}, {Fern{\'a}ndez-Trincado},
  {Frinchaboy}, {Hearty}, \& {Nitschelm}}]{Pinsonneault2018}
{Pinsonneault}, M.~H., {Elsworth}, Y.~P., {Tayar}, J., {et~al.} 2018, \apjs,
  239, 32

\bibitem[{{Pont} \& {Eyer}(2004)}]{pont04}
{Pont}, F., \& {Eyer}, L. 2004, \mnras, 351, 487

\bibitem[{{Pu} \& {Lai}(2019)}]{Pu2019}
{Pu}, B., \& {Lai}, D. 2019, \mnras, 488, 3568

\bibitem[{{Raghavan} {et~al.}(2010){Raghavan}, {McAlister}, {Henry}, {Latham},
  {Marcy}, {Mason}, {Gies}, {White}, \& {ten Brummelaar}}]{Raghavan2010}
{Raghavan}, D., {McAlister}, H.~A., {Henry}, T.~J., {et~al.} 2010, \apjs, 190,
  1

\bibitem[{{Ricker} {et~al.}(2014){Ricker}, {Winn}, {Vanderspek}, {Latham},
  {Bakos}, {Bean}, {Berta-Thompson}, {Brown}, {Buchhave}, {Butler}, {Butler},
  {Chaplin}, {Charbonneau}, {Christensen-Dalsgaard}, {Clampin}, {Deming},
  {Doty}, {De Lee}, {Dressing}, {Dunham}, {Endl}, {Fressin}, {Ge}, {Henning},
  {Holman}, {Howard}, {Ida}, {Jenkins}, {Jernigan}, {Johnson}, {Kaltenegger},
  {Kawai}, {Kjeldsen}, {Laughlin}, {Levine}, {Lin}, {Lissauer}, {MacQueen},
  {Marcy}, {McCullough}, {Morton}, {Narita}, {Paegert}, {Palle}, {Pepe},
  {Pepper}, {Quirrenbach}, {Rinehart}, {Sasselov}, {Sato}, {Seager},
  {Sozzetti}, {Stassun}, {Sullivan}, {Szentgyorgyi}, {Torres}, {Udry}, \&
  {Villasenor}}]{ricker14}
{Ricker}, G.~R., {Winn}, J.~N., {Vanderspek}, R., {et~al.} 2014, in Society of
  Photo-Optical Instrumentation Engineers (SPIE) Conference Series, Vol. 9143,
  , 20

\bibitem[{{Rizzuto}(2017)}]{RizzutoProp}
{Rizzuto}, A. 2017, {The Exoplanet Migration Timescale from K2 Young Clusters},
  NASA ADAP Proposal, ,

\bibitem[{{Rizzuto} {et~al.}(2018){Rizzuto}, {Vanderburg}, {Mann}, {Kraus},
  {Dressing}, {Ag{\"u}eros}, {Douglas}, \& {Krolikowski}}]{Rizzuto2018}
{Rizzuto}, A.~C., {Vanderburg}, A., {Mann}, A.~W., {et~al.} 2018, \aj, 156, 195

\bibitem[{{Schlichting} \& {Chang}(2011)}]{Schlichting2011}
{Schlichting}, H.~E., \& {Chang}, P. 2011, \apj, 734, 117

\bibitem[{Scholz \& Stephens(1987)}]{Scholz1987}
Scholz, F.~W., \& Stephens, M.~A. 1987, Journal of the American Statistical
  Association, 82, 918.
\newblock \url{https://doi.org/10.1080/01621459.1987.10478517}

\bibitem[{{Scott}(1992)}]{Scott1992}
{Scott}, D.~W. 1992, {Multivariate Density Estimation}

\bibitem[{{Seager} \& {Mall{\'e}n-Ornelas}(2003)}]{seager03}
{Seager}, S., \& {Mall{\'e}n-Ornelas}, G. 2003, \apj, 585, 1038

\bibitem[{{Silva Aguirre} {et~al.}(2015){Silva Aguirre}, {Davies}, {Basu},
  {Christensen-Dalsgaard}, {Creevey}, {Metcalfe}, {Bedding}, {Casagrande},
  {Handberg}, {Lund}, {Nissen}, {Chaplin}, {Huber}, {Serenelli}, {Stello}, {Van
  Eylen}, {Campante}, {Elsworth}, {Gilliland}, {Hekker}, {Karoff}, {Kawaler},
  {Kjeldsen}, \& {Lundkvist}}]{Aguirre2015}
{Silva Aguirre}, V., {Davies}, G.~R., {Basu}, S., {et~al.} 2015, \mnras, 452,
  2127

\bibitem[{{Skumanich}(1972)}]{Skumanich1972}
{Skumanich}, A. 1972, \apj, 171, 565

\bibitem[{{Sliski} \& {Kipping}(2014)}]{sliski14}
{Sliski}, D.~H., \& {Kipping}, D.~M. 2014, \apj, 788, 148

\bibitem[{{Soderblom}(2010)}]{Soderblom2010}
{Soderblom}, D.~R. 2010, \araa, 48, 581

\bibitem[{{Spalding} \& {Batygin}(2016)}]{Spalding2016}
{Spalding}, C., \& {Batygin}, K. 2016, \apj, 830, 5

\bibitem[{Taillon(2018)}]{Gabinou2018}
Taillon, G. 2018, KS2D, \url{https://github.com/Gabinou/2DKS},  GitHub

\bibitem[{Tange(2018)}]{Tange2018}
Tange, O. 2018, GNU Parallel 2018 (Ole Tange), doi:10.5281/zenodo.1146014.
\newblock \url{https://doi.org/10.5281/zenodo.1146014}

\bibitem[{{Thompson} {et~al.}(2018){Thompson}, {Coughlin}, {Hoffman},
  {Mullally}, {Christiansen}, {Burke}, {Bryson}, {Batalha}, {Haas},
  {Catanzarite}, {Rowe}, {Barentsen}, {Caldwell}, {Clarke}, {Jenkins}, {Li},
  {Latham}, {Lissauer}, {Mathur}, {Morris}, {Seader}, {Smith}, {Klaus},
  {Twicken}, {Van Cleve}, {Wohler}, {Akeson}, {Ciardi}, {Cochran}, {Henze},
  {Howell}, {Huber}, {Pr{\v s}a}, {Ram{\'{\i}}rez}, {Morton}, {Barclay},
  {Campbell}, {Chaplin}, {Charbonneau}, {Christensen-Dalsgaard}, {Dotson},
  {Doyle}, {Dunham}, {Dupree}, {Ford}, {Geary}, {Girouard}, {Isaacson},
  {Kjeldsen}, {Quintana}, {Ragozzine}, {Shabram}, {Shporer}, {Silva Aguirre},
  {Steffen}, {Still}, {Tenenbaum}, {Welsh}, {Wolfgang}, {Zamudio}, {Koch}, \&
  {Borucki}}]{Thompson2018}
{Thompson}, S.~E., {Coughlin}, J.~L., {Hoffman}, K., {et~al.} 2018, \apjs, 235,
  38

\bibitem[{{Thorngren} {et~al.}(2016){Thorngren}, {Fortney}, {Murray-Clay}, \&
  {Lopez}}]{Thorngren2016}
{Thorngren}, D.~P., {Fortney}, J.~J., {Murray-Clay}, R.~A., \& {Lopez}, E.~D.
  2016, \apj, 831, 64

\bibitem[{van~der Walt {et~al.}(2014)van~der Walt, {S}ch\"onberger,
  {Nunez-Iglesias}, {B}oulogne, {W}arner, {Y}ager, {G}ouillart, {Y}u, \& the
  scikit-image contributors}]{skimage}
van~der Walt, S., {S}ch\"onberger, J.~L., {Nunez-Iglesias}, J., {et~al.} 2014,
  PeerJ, 2, e453.
\newblock \url{https://doi.org/10.7717/peerj.453}

\bibitem[{{Van Eylen} {et~al.}(2018){Van Eylen}, {Agentoft}, {Lundkvist},
  {Kjeldsen}, {Owen}, {Fulton}, {Petigura}, \& {Snellen}}]{VanEylen2018}
{Van Eylen}, V., {Agentoft}, C., {Lundkvist}, M.~S., {et~al.} 2018, \mnras,
  479, 4786

\bibitem[{{Van Eylen} {et~al.}(2019){Van Eylen}, {Albrecht}, {Huang},
  {MacDonald}, {Dawson}, {Cai}, {Foreman-Mackey}, {Lundkvist}, {Silva Aguirre},
  {Snellen}, \& {Winn}}]{VanEylen2019}
{Van Eylen}, V., {Albrecht}, S., {Huang}, X., {et~al.} 2019, \aj, 157, 61

\bibitem[{{Verner} {et~al.}(2011){Verner}, {Elsworth}, {Chaplin}, {Campante},
  {Corsaro}, {Gaulme}, {Hekker}, {Huber}, {Karoff}, {Mathur}, {Mosser},
  {Appourchaux}, {Ballot}, {Bedding}, {Bonanno}, {Broomhall}, {Garc{\'{\i}}a},
  {Handberg}, {New}, {Stello}, {R{\'e}gulo}, {Roxburgh}, {Salabert}, {White},
  {Caldwell}, {Christiansen}, \& {Fanelli}}]{verner11}
{Verner}, G.~A., {Elsworth}, Y., {Chaplin}, W.~J., {et~al.} 2011, \mnras, 415,
  3539

\bibitem[{{Vick} {et~al.}(2019){Vick}, {Lai}, \& {Anderson}}]{Vick2019}
{Vick}, M., {Lai}, D., \& {Anderson}, K.~R. 2019, \mnras, 484, 5645

\bibitem[{{Virtanen} {et~al.}(2020){Virtanen}, {Gommers}, {Oliphant},
  {Haberland}, {Reddy}, {Cournapeau}, {Burovski}, {Peterson}, {Weckesser},
  {Bright}, {van der Walt}, {Brett}, {Wilson}, {Jarrod Millman}, {Mayorov},
  {Nelson}, {Jones}, {Kern}, {Larson}, {Carey}, {Polat}, {Feng}, {Moore}, {Vand
  erPlas}, {Laxalde}, {Perktold}, {Cimrman}, {Henriksen}, {Quintero}, {Harris},
  {Archibald}, {Ribeiro}, {Pedregosa}, {van Mulbregt}, \&
  {Contributors}}]{Scipy}
{Virtanen}, P., {Gommers}, R., {Oliphant}, T.~E., {et~al.} 2020, Nature
  Methods, 17, 261

\bibitem[{{Weiss} {et~al.}(2018){Weiss}, {Isaacson}, {Marcy}, {Howard},
  {Petigura}, {Fulton}, {Winn}, {Hirsch}, {Sinukoff}, {Rowe}, \& {California
  Kepler Survey}}]{Weiss2018b}
{Weiss}, L.~M., {Isaacson}, H.~T., {Marcy}, G.~W., {et~al.} 2018, \aj, 156, 254

\bibitem[{{Wu}(2019)}]{Wu2019}
{Wu}, Y. 2019, \apj, 874, 91

\bibitem[{{Xie} {et~al.}(2016){Xie}, {Dong}, {Zhu}, {Huber}, {Zheng}, {De Cat},
  {Fu}, {Liu}, {Luo}, {Wu}, {Zhang}, {Zhang}, {Zhou}, {Cao}, {Hou}, {Wang}, \&
  {Zhang}}]{xie16}
{Xie}, J.-W., {Dong}, S., {Zhu}, Z., {et~al.} 2016, Proceedings of the National
  Academy of Science, 113, 11431

\end{thebibliography}

\end{document}